\documentstyle[amstex,aps]{revtex}

\begin{document}
\title{The Algebraic Model for scattering in three-s-cluster systems.\\
I. Theoretical Background}
\author{V. Vasilevsky, A. V. Nesterov, F. Arickx, J. Broeckhove}
\address{Bogolyubov Institute for Theoretical Physics, Kiev, Ukraine\\
Universiteit Antwerpen (RUCA), Dept. of Mathematics and Computer Science,\\
Antwerp, Belgium}
\date{\today}
\maketitle

\begin{abstract}
A framework to calculate two-particle matrix elements for fully
antisymmetrized three-cluster configurations is presented. The theory is
developed for a scattering situation described in terms of the Algebraic
Model. This means that the nuclear many-particle state and its asymptotic
behaviour are expanded in terms of oscillator states of the intra-cluster
coordinates. The Generating Function technique is used to optimize the
calculation of matrix elements. In order to derive the dynamical equations,
a multichannel version of the Algebraic Model is presented.
\end{abstract}

\section{Introduction}

Since 1980, the so-called Algebraic Model (AM) of the Resonating Group
Method has been used in the investigation of bound and continuum states of
nuclear systems. Initially the AM was applied to binary cluster
configurations \cite
{kn:Fil_Okhr,kn:Fil81,vasil_rybk89,mirror_react90,kn:fil_rev2}. Later on it
was extended to describe binary clusters coupled to collective (quadrupole
and monopole) channels \cite{kn:fil_rev2,kn:fil_vasil85,nest_8be,nest_psh}.
Quite recently, three-cluster configurations were considered in the AM
framework \cite{kn:Vasil96,kn:Vasil97,kn:Fil_rev4}. Such configurations play
a significant role in light nuclei, in particular for reactions of
astrophysical interest.

The Algebraic Model represents the nuclear many-particle wave functions
through their expansions in harmonic oscillator eigenstates. The use of a
basis of square integrable states reduces the Schr\"{o}dinger equation to a
matrix equation. This procedure is well-known for bound states, but is also
applicable to continuum states when the appropriate boundary conditions are
imposed on the expansion coefficients \cite
{kn:Fil_Okhr,kn:Fil81,kn:Heller1,kn:Heller2,kn:Smirnov}. Thus the Algebraic
Model provides a unified approach to bound and continuous spectra based on
familiar matrix techniques.

Solving an AM\ scattering problem requires two major steps. The first is to
define the basis states relevant to the scattering channels being
investigated and to compute the Hamiltonian matrix elements in that basis.
The second is to consider the description of the asymptotic region and
boundary conditions in the basis and to solve the matrix equation subject to
those boundary conditions.

We elaborate on both steps in this paper for a fully antisymmetrized
three-cluster system, and introduce the proper three-cluster continuum
boundary conditions. Such a description is still lacking in the literature.
Three-cluster systems have been considered in some calculations, but there
the continuum was approximated through sets of two-cluster configurations 
\cite{kn:Desc94,kn:Varga+96,kn:Csoto+Lang99}.

In our calculation the Pauli principle will be treated rigorously by taking
full antisymmetrization into account for all nucleons in the system. The
basis states will be defined by a composition of three s-clusters with
``frozen'' internal structure. The relative motions of the clusters are the
relevant degrees of freedom. They are described by the Hyperspherical
Harmonics method (HHM). It makes a natural formulation of the proper
boundary conditions possible. The matrix elements of kinetic and potential
energy in the three-cluster basis states are calculated using the Generating
Function technique \cite{kn:fil_rev2,kn:fil_rev1,kn:fil_rev3}, and methods
for an explicit derivation will be presented.

The asymptotic behavior of relative motion of clusters is obtained by
considering large inter-cluster distances where the interaction between
clusters approaches zero. We use the folding approximation to determine the
asymptotic wave-functions. The folding approximation amounts to solving the
scattering equations without nucleon-nucleon interaction and without
inter-cluster antisymmetrization.

In the case of neutral clusters, i.e. when inter-cluster Coulomb
interactions are absent, the asymptotic equations are uncoupled in the
hyperangular momentum $K$ that is typical in the HHM. We will therefore
introduce individual scattering channels characterized by quantum number $K$%
. In configurations with charged clusters the Coulomb interaction also
couples $K$-channels in the asymptotic region. Again the folding
approximation will be used to arrive at the appropriate asymptotic wave
equations. A multiple channel approach which couples all $K$-channels has to
be used to solve the full AM system of equations. This extends the
theoretical foundation of the AM given in \cite{kn:VA_PR}\ for two-cluster
systems.

In part II of this paper, the techniques discussed here will be applied to
the three cluster configurations $\alpha +n+n$ for $^{6}$He, and $\alpha
+p+p $ for $^{6}$Be.

\section{Three-s-cluster systems}

\subsection{Wave Functions\label{sec:WaveFions}}

The many-particle wave function for a three-cluster system of $A$ nucleons ($%
A=A_{1}+A_{2}+A_{3}$) can be written, using the anti-symmetrization operator 
${\cal A}$, as follows

\begin{equation}
\Psi \left( {\bf q}_{1},..,{\bf q}_{A-1}\right) ={\cal A}\left[ \Psi
_{1}\left( A_{1}\right) \ \Psi _{2}\left( A_{2}\right) \ \Psi _{3}\left(
A_{3}\right) \ \Psi _{R}\left( R\right) \right]  \label{eq:AntisymClustState}
\end{equation}
where the centre of mass of the $A$-nucleon system has been eliminated by
the use of Jacobi coordinates ${\bf q}_{i}$. The cluster wave functions $%
\Psi _{i}\left( A_{i}\right) $ 
\begin{equation}
\Psi _{i}\left( A_{i}\right) =\Psi _{i}\left( {\bf q}_{1}^{\left( i\right)
},..,{\bf q}_{A_{i}-1}^{\left( i\right) }\right) \quad (i=1,2,3)
\label{eq:ClustState}
\end{equation}
represent the internal structure of the $i$-th cluster, centered around its
centre of mass ${\bf R}_{i}$. In our AM study these cluster functions are
fixed and they are Slater determinants of harmonic oscillator (0s)-states,
corresponding to the groundstate configuration of the cluster ($A_{i}\leq 4$
for all $i$). The $\Psi _{R}\left( R\right) $ wave function 
\begin{equation}
\Psi _{R}\left( R\right) =\Psi _{R}\left( {\bf q}_{1}^{\left( R\right) },%
{\bf q}_{2}^{\left( R\right) }\right) =\Psi _{R}\left( {\bf q}_{1},{\bf q}%
_{2}\right)  \label{eq:RelMotionState}
\end{equation}
represents the relative motion of the three clusters with respect to one
another, and ${\bf q}_{1}$ and ${\bf q}_{2}$ represent Jacobi coordinates.
In Fig. \ref{fig:figure1} we indicate an enumeration of possible Jacobi
coordinates and their relation to the component clusters.

\begin{figure}[tbp]
\caption{Two configurations of the Jacobi coordinates for the three-cluster
system $\protect\alpha +N+N$.}
\label{fig:figure1}
\end{figure}

The state (\ref{eq:RelMotionState}) is not limited to any particular type of
orbital; on the contrary we will use a complete basis of harmonic oscillator
states for the relative motion degrees of freedom. Thus the full $A$%
-particle state cannot be expressed as a single Slater determinant of single
particle orbitals.

An important approximation is obtained by breaking the Pauli principle
between the individual clusters, but retaining a proper quantum-mechanical
description of the clusters, which is described by the wave function 
\begin{equation}
\Psi _{F}\left( {\bf q}_{1},..,{\bf q}_{A-1}\right) =\Psi _{1}\left(
A_{1}\right) \ \Psi _{2}\left( A_{2}\right) \ \Psi _{3}\left( A_{3}\right) \
\Psi _{R}\left( R\right)
\end{equation}
Because each cluster wave function is antisymmetric (they are Slater
determinants) one is indeed neglecting the inter-cluster anti-symmetrization
only. This results in what is known as the ``Folding'' model. It has the
advantage of preserving the identities of the clusters and, if the
intra-cluster structure is kept ``frozen'', it reduces the many-particle
problem to that of the relative motion of the clusters.

The folding approximation will be the natural choice for calculating the
asymptotic behavior of the three cluster-system, i.e. the disintegration of
the system in the three non-interacting individual clusters. This amounts to
the situation that all three clusters are a sufficient distance apart and
intra-cluster interactions are no longer in force.

The folding model is however also an acceptable approximation in the
interaction region and can serve as comparison to the fully antisymmetrized
calculations. In the folding model, the clusters interact through a local,
inter-cluster potential called the folding potential. As in the current
paper (0s)-determinants $\Psi _{i}\left( A_{i}\right) $ are used to describe
the internal state of the clusters, the folding potential will be easily
calculated and is a sum of three terms 
\begin{equation}
V^{(F)}=V^{(F)}({\bf R}_{12})+V^{(F)}({\bf R}_{23})+V^{(F)}({\bf R}_{31})
\label{eq:FoldPot}
\end{equation}
where each term is simply the integral 
\begin{equation}
V^{(F)}({\bf R}_{\tau \upsilon })=\sum_{i\in A_{\tau }}\sum_{j\in
A_{\upsilon }}\int d\tau _{\tau }d\tau _{\upsilon }|\Psi _{\tau }(A_{\tau
})|^{2\ }V({\bf r}_{i}-{\bf r}_{j}+{\bf R}_{\tau \upsilon })|\ \Psi
_{\upsilon }(A_{\upsilon })|^{2}  \label{eq:FoldPotTerm}
\end{equation}
The coordinates ${\bf R}_{\tau \upsilon }$ are associated with the relative
position of the clusters 
\begin{equation}
{\bf R}_{\tau \upsilon }=\frac{1}{A_{\tau }}\sum_{i\in A_{\tau }}{\bf r}_{i}-%
\frac{1}{A_{\upsilon }}\sum_{j\in A_{\upsilon }}{\bf r}_{j}
\label{eq:RelPosClust}
\end{equation}
and sum to zero; they are equivalent to the ${\bf q}_{1},{\bf q}_{2}$ Jacobi
coordinates introduced earlier. In this way the folding approximation turns
the three-cluster problem into an effective three-particle problem for the
relative motion coordinates.

Because the cluster states are fixed and built up of (0s)-orbitals, the
problem of labeling the basis states with quantum numbers relates to the
inter-cluster wave function only. This holds true whether one uses the full
anti-symmetrization or the folding approximation. In a two-cluster case, the
set of quantum numbers describing inter-cluster motion is unambiguously
defined. In a three-cluster case, several schemes can be used to classify
the inter-cluster wave function in the oscillator representation. In \cite
{kn:Vasil96,kn:Vasil97} three distinct but equivalent schemes were
considered. One of these used the quantum numbers provided by the
Hyperspherical Harmonics (HH) method (see for instance \cite
{kn:Simon68,kn:Fabr93,kn:Zhuk93}). This is the classification that we will
adopt. Even within this particular scheme there are several ways to classify
the basis states. We shall restrict ourselves to the so-called
Zernike-Brinkman basis \cite{kn:ZB}. This corresponds to the following
reduction of the unitary group $U(6)$, the symmetry group of the
three-particle oscillator Hamiltonian, 
\begin{equation}
U(6)\supset O(6)\supset O(3)\otimes O(3)\supset O(3)  \label{eq:GroupReduc}
\end{equation}
This reduction provides the quantum numbers $K$, the hypermomentum, $n$, the
hyperradial excitation, $l_{1}$, the angular momentum connected with the
first Jacobi vector, $l_{2}$, the angular momentum connected with the second
Jacobi vector, and $L$ and $M$ the total angular momentum obtained from
coupling the partial angular momenta $l_{1}$, $l_{2}$. Collectively these
quantum numbers will be denoted by $\nu $, i.e. $\nu =\{n,K,(l_{1}l_{2})LM\}$
in the remainder of the text.

There are a number of relations and constraints on these quantum numbers:

\begin{itemize}
\item  the total angular momentum is the vector sum of the partial angular
momenta ${\bf l}_{1}$and ${\bf l}_{2}$, i.e. ${\bf L}=$ ${\bf l}_{1}+$ ${\bf %
l}_{2}$ or $\left| l_{1}-l_{2}\right| \leq L\leq l_{1}+l_{2}$.

\item  by fixing the values of $l_{1}$ and $l_{2}$, we impose restrictions
on the hypermomentum $K=l_{1}+l_{2},~l_{1}+l_{2}+2,~l_{1}+l_{2}+4,\ldots $
This condition implies that for certain values of hypermomentum $K$ the sum
of partial angular momenta $l_{1}+l_{2}$ cannot exceed $K$.

\item  the partial angular momenta $l_{1}$ and $l_{2}$ define the parity of
the three-cluster state by the relation $\pi =\left( -1\right)
^{l_{1}+l_{2}} $.

\item  for the ``normal'' parity states $\pi =\left( -1\right) ^{L}$ the
minimal value of hypermomentum is $K_{\min }=L$, whereas $K_{\min }=L+1$ for
the so-called ``abnormal'' parity states $\pi =\left( -1\right) ^{L+1}$.

\item  oscillator shells with $N$ quanta are characterized by the constraint 
$N=2n\;+K$.
\end{itemize}

Thus for a given hyperangular and rotational configuration the quantumnumber 
$n$ ladders the oscillator shells of increasing oscillator energy.

\section{The Algebraic Model}

\subsection{Asymptotic solutions in coordinate representation}

The Algebraic Model implements a method to solve the Schr\"{o}dinger
equation for quantum scattering systems, in particular for nuclear cluster
systems. It is based on a matrix representation of the Schr\"{o}dinger
equation in terms of a square integrable basis, usually Harmonic Oscillator
states, and boundary conditions in terms of the asymptotic behavior of the
expansion coefficients of the wave function. In this paper we restrict
ourselves to a presentation tailored to the treatment of three-cluster
systems.

In the case of three-cluster calculations, one needs to determine a proper
approximation for the wave function (\ref{eq:RelMotionState}). Consider an
expansion of the relative wave function 
\begin{equation}
\Psi _{R}\left( {\bf q}_{1},{\bf q}_{2}\right) =\sum_{\nu }c_{\nu }\Psi
_{\nu }\left( {\bf q}_{1},{\bf q}_{2}\right)  \label{eq:RelWaveFuncAsympt}
\end{equation}
with $\nu =\{n,K,(l_{1}l_{2})LM\}$ and $\left\{ \Psi _{\nu }\right\} $ a
complete basis of six-dimensional oscillator states. It covers all possible
types of relative motion between the three clusters.

To obtain the asymptotic behavior of the three-cluster system, we consider
the folding approximation. The assumption that antisymmetrization effects
between clusters are absent in the asymtptotic region is a natural one. The
relative motion problem of the three clusters in the absence of a potential
can then be explicitly solved in the Hyperspherical Harmonics (HH) method
(see for instance \cite{kn:Simon68,kn:Fabr93,kn:Zhuk93}). It involves the
transformation of the Jacobi coordinates ${\bf q}_{1}$ and ${\bf q}_{2}$\ to
the hyperradius $\rho $\ and a set of hyperangles $\Omega $. The
inter-cluster wave function in coordinate representation is expanded in
hyperspherical harmonics $H_{K}^{\nu _{0}}\left( \Omega \right) $ where $\nu
_{0}$ has been chosen as a shorthand for $(l_{1}l_{2})LM$.

In the absence of the Coulomb interaction this leads to a set of equations
for the hyperradial asymptotic solutions, with the kinetic energy operator
as reference Hamiltonian 
\begin{equation}
\left\{ -\frac{\hbar ^{2}}{2m}\left[ \frac{d^{2}}{d\rho ^{2}}+\frac{5}{\rho }%
\frac{d}{d\rho }-\frac{K\left( K+4\right) }{\rho ^{2}}\right] -E\right\}
R_{K,\nu _{0}}\left( \rho \right) =0
\end{equation}

The solutions can be obtained analytically and are represented by a pair of
H\"{a}nkel functions for the ingoing and outgoing solutions: 
\begin{equation}
R_{K,\nu _{0}}^{\left( \pm \right) }\left( \rho \right) =\left\{ 
\begin{array}{l}
H_{K+2}^{\left( 1\right) }\left( k\rho \right) /\rho ^{2} \\ 
H_{K+2}^{\left( 2\right) }\left( k\rho \right) /\rho ^{2}
\end{array}
\right\}
\end{equation}
where 
\[
k=\sqrt{\frac{2mE}{\hbar ^{2}}} 
\]
One notices that these asymptotic solutions are independent of all quantum
numbers $\nu _{0}$, and are determined by the value of hypermomentum $K$
only.

When charged clusters are considered the asymptotic reference Hamiltonian
consists of the kinetic energy and the Coulomb interaction:

\begin{equation}
\left\{ -\frac{\hbar ^{2}}{2m}\left[ \frac{d^{2}}{d\rho ^{2}}+\frac{5}{\rho }%
\frac{d}{d\rho }-\frac{\left\| {\cal K}\right\| }{\rho ^{2}}\right] +\frac{%
\left\| Z_{eff}\right\| }{\rho }-E\right\} \left\| {\cal R}\left( \rho
\right) \right\| =0
\end{equation}
The matrix $\left\| {\cal K}\right\| $ is diagonal with matrix elements $%
K\left( K+4\right) $, and $\left\| Z_{eff}\right\| $, the ``effective
charge'', is off-diagonal in $K$ and $(l_{1}l_{2})$. Different $K$-channels
are now coupled. A standard approximation for solving these equations is to
decouple them by assuming that the off-diagonal matrix-elements of $\left\|
Z_{eff}\right\| $ are sufficiently small: 
\begin{equation}
\left\{ -\frac{\hbar ^{2}}{2m}\left[ \frac{d^{2}}{d\rho ^{2}}+\frac{5}{\rho }%
\frac{d}{d\rho }-\frac{K\left( K+4\right) }{\rho ^{2}}\right] +\frac{Z_{eff}%
}{\rho }-E\right\} R_{K,\nu _{0}}\left( \rho \right) =0
\end{equation}
The constants $Z_{eff}$ depends on $K$ and $\nu _{0}$ and all parameters of
the many-body system under consideration. We will restrict ourselves to this
decoupling approximation, but it is to be understood that its validity has
to be checked for any specific three-cluster system.

The asymptotic solutions then become 
\begin{equation}
R_{K,\nu _{0}}^{\left( \pm \right) }\left( \rho \right) =\left\{ 
\begin{array}{l}
W_{i\eta ,\mu }\left( 2ik\rho \right) /\rho ^{\frac{5}{2}} \\ 
W_{-i\eta ,\mu }\left( -2ik\rho \right) /\rho ^{\frac{5}{2}}
\end{array}
\right\}
\end{equation}
where $W$ is the Whittaker function, $\mu =K+2$ and $\eta $\ is the
well-known Sommerfeld parameter 
\begin{equation}
\eta =\frac{m}{\hbar ^{2}}\frac{Z_{eff}}{k}
\end{equation}

As $\eta $\ is a function of $K,$ $l_{1}$ and $l_{2}$ through the parameter $%
Z_{eff}$, the asymptotic solutions will now be dependent on $K$ and $\nu
_{0} $.

\subsection{Asymptotic solutions in oscillator representation}

The Algebraic Model relies on an expansion in terms of oscillator functions,
and the asymptotic behavior of the corresponding expansion coefficients $%
c_{\nu }$. It was conjectured (see for instance \cite{kn:VA_PR}) that for
very large values of the oscillator quantum number $n$ the expansion
coefficients for physically relevant wave-functions behave like 
\begin{equation}
c_{n}=\left\langle n|\psi \right\rangle \simeq \sqrt{2}\rho _{n}^{2}\psi
(b\rho _{n})  \label{eq:AsymptoticForCnGeneral}
\end{equation}
where $\rho _{n}=\sqrt{4n+2K+6}$ corresponds to the classical turning point, 
$b$ is the oscillator parameter, and $\psi $ is the hyperradial wave
function.

In the case of neutral clusters this leads after substitution of the
hyperradial asymptotic solutions to the following expansion coefficients $%
c_{n}^{\left( \pm \right) }$ 
\begin{equation}
c_{n}^{\left( \pm \right) K}\simeq \sqrt{2}\left\{ 
\begin{array}{l}
H_{K+2}^{\left( 1\right) }\left( kb\rho _{n}\right) \\ 
H_{K+2}^{\left( 2\right) }\left( kb\rho _{n}\right)
\end{array}
\right\}  \label{eq:AsymCoefFreePart}
\end{equation}

This result can be obtained in an alternative way \cite{kn:4n,kn:AM_12C} by
representing the Schr\"{o}dinger equation, with the kinetic energy operator $%
\hat{T}$ as the Hamiltonian to describe the asymptotic situation, in a
(hyperradial) oscillator representation 
\begin{equation}
\sum_{m=0}^{\infty }\left\langle n,\left( K,\nu _{0}\right) \left| \hat{T}%
-E\right| m,\left( K,\nu _{0}\right) \right\rangle c_{m}^{K,\nu _{0}}=0
\end{equation}
This matrix equation is of a three-diagonal form because of the properties
of $\hat{T}$ and the oscillator basis. Solving for the expansion
coefficients $c_{n}^{K,\nu _{0}}$ leads to a three-term recurrence relation 
\begin{equation}
T_{n,n-1}^{K,\nu _{0}}c_{n-1}^{K,\nu _{0}}+\left( T_{n,n}^{K,\nu
_{0}}-E\right) c_{n}^{K,\nu _{0}}+T_{n,n+1}^{K,\nu _{0}}c_{n+1}^{K,\nu
_{0}}=0
\end{equation}
where 
\begin{equation}
T_{n,m}^{K,\nu _{0}}=\left\langle n,\left( K,\nu _{0}\right) \left| \hat{T}%
\right| m,\left( K,\nu _{0}\right) \right\rangle
\end{equation}
The asymptotic solutions (i.e. for high $n$) of this recurrence relation are
then precisely given by (\ref{eq:AsymCoefFreePart}).

When the Coulomb interaction is present we again apply (\ref
{eq:AsymptoticForCnGeneral}) to obtain 
\begin{equation}
c_{n}^{\left( \pm \right) K}\simeq \sqrt{2}\left\{ 
\begin{array}{l}
W_{i\eta ,\mu }\left( 2ikb\rho _{n}\right) /\sqrt{\rho _{n}} \\ 
W_{-i\eta ,\mu }\left( -2ikb\rho _{n}\right) /\sqrt{\rho _{n}}
\end{array}
\right\}  \label{eq:AsymCoefCoulPart}
\end{equation}
In this case the oscillator representation of the Schr\"{o}dinger equation
is no longer of a tridiagonal form, and cannot be solved analytically for
the asymptotic solutions to corroborate this result.

It should be noted, that the above elaborations are valid for relatively
small values of momentum $k$ and sufficiently large values of discrete
hyperradius $\rho _{n}$, or when 
\begin{equation}
\frac{k^{2}}{\rho _{n}^{2}}\ll 1
\end{equation}
which defines the ``asymptotic regime''; it shows that for any value of $k$
one can find values for $n$ where the relation is satisfied.

As we consider an asymptotic decoupling in the $\left( K,\nu _{0}\right) $
quantumnumbers, one will deal with asymptotic channels characterized by the $%
\left( K,\nu _{0}\right) $ values. So only in the internal (or interaction)
region will states with different $K$ and $\nu _{0}$ be coupled by the
short-range nuclear potential and the Coulomb potential. The three-cluster
system can therefore be described by a coupled-channels approach, where the
individual channels are characterized by a single $K$-value, and we will
henceforth refer to these channels as ``$K$-channels''.

\subsection{Multi-channel AM equations}

In the current many-channel description of the Algebraic Model for
three-cluster systems, the channels will be characterized by the a specific
value of the set of quantumnumbers $K,\nu _{0}$, whereas the relative motion
of clusters within the channel is connected to the oscillator index $n$. We
will use $K$ henceforth as a corporate index for individual channels, and
assume it represents all $K,\nu _{0}$ quantum numbers.

The Schr\"{o}dinger equation can be cast in a matrix equation of the form 
\begin{equation}
\sum_{K^{\prime }}\sum_{m}\left\langle n,K\left| \hat{H}-E\right|
m,K^{\prime }\right\rangle \ c_{m}^{K^{\prime }}=0  \label{eq:CouplChanEqs1}
\end{equation}

We will now use a representation of the dynamical equations presented in 
\cite{kn:VA_PR}. As we will consider an $S$-matrix formulation of the
problem, the expansion coefficients are rewritten as 
\begin{equation}
c_{n}^{K}=c_{n}^{(0)K}+\delta _{K_{i}K}c_{n}^{(-)K}-S_{K_{i}K}c_{n}^{(+)K}
\label{eq:AsymptExpOfCn}
\end{equation}
where, for the current channel $K$, the $c_{n}^{(0)K}$\ are the so-called
residual coefficients, the $c_{n}^{(\pm )K}$ are the incoming and outgoing
asymptotic coefficients (valid for all $n$). The matrix element $S_{K_{i}K}$
describes the coupling between the current channel $K$ and the entrance
channel $K_{i}$.

As shown in \cite{kn:Heller1,kn:Yamani,kn:Smirnov}, the $c_{n}^{(\pm )K}$
satisfy the following system of equations for a given channel $K$%
\begin{equation}
\sum_{m=0}^{\infty }\left\langle n,K\left| \hat{H}_{0}-E\right|
m,K\right\rangle \ c_{m}^{(\pm )K}=\beta _{0}^{(\pm )K}\ \delta _{n,0}
\label{eq:AsymptEqnsKchannels}
\end{equation}
$\hat{H}_{0}$ being the asymptotic reference Hamiltonian, which equals the
kinetic energy operator for uncharged clusters, and the kinetic energy
operator plus Coulomb interaction for charged clusters. The right-hand side
features $\beta _{0}^{(\pm )K}$ which is a regularization factor to account
for the irregular behavior of the $c_{0}^{(\pm )K}$. This factor allows one
to solve (\ref{eq:AsymptEqnsKchannels}) for all values of $n$. The value of $%
\beta _{0}^{(\pm )K}$ can be obtained for both reference Hamiltonians (i.e.
with or without Coulomb). The set of equations (\ref{eq:AsymptEqnsKchannels}%
)\ for the asymptotic coefficients can then be solved numerically to
different degrees of approximation depending on the requested precision. The 
$c_{n}^{K\left( \pm \right) \text{ }}$have the desired asymptotic behavior
(cfr eqs \ref{eq:AsymCoefFreePart} and \ref{eq:AsymCoefCoulPart})

Substitution of (\ref{eq:AsymptExpOfCn}) in the equations (\ref
{eq:CouplChanEqs1}) then leads to the following system of dynamical
equations for the many-channel system: 
\begin{eqnarray}
\sum_{K^{\prime }}\sum_{m}\left\langle n,K\left| \hat{H}-E\right|
m,K^{\prime }\right\rangle \ c_{m}^{K^{\prime }\left( 0\right) }-
&&\sum_{K^{\prime }}S_{K_{i}K^{\prime }}\left[ \beta _{0}^{(+)K^{\prime }}\
\delta _{n,0}\ \delta _{K^{\prime }K}+V_{n}^{KK^{\prime }\left( +\right) }%
\right]  \nonumber \\
=- &&\beta _{0}^{(-)K}\ \delta _{n,0}\ \delta _{K_{i}K}-V_{n}^{KK_{i}\left(
-\right) }  \label{eq:CouplChanEqs2}
\end{eqnarray}
where the dynamical coefficients $V_{n}^{KK^{\prime }\left( \pm \right) }$,
defined in \cite{kn:VA_PR}, are given by 
\begin{equation}
V_{n}^{(\pm )KK^{\prime }}=\sum_{m=0}^{\infty }\left\langle n,K\left| \hat{V}%
\right| m,K^{\prime }\right\rangle \ c_{m}^{(\pm )K^{\prime }}
\end{equation}

This system of equations should be solved for both the residual coefficients 
$c_{n}^{K\left( 0\right) }$ and the $S$-matrix elements $S_{K^{\prime }K}$.

To obtain an appropriate approximation to the exact solution of (\ref
{eq:CouplChanEqs2}), we consider an internal region corresponding to $n<N$
and an asymptotic region with $n\geq N$. The choice of $N$ is such that one
can expect the residual expansion coefficients $\left\{ c_{n}^{(0)K}\right\} 
$ to be sufficiently small in the asymptotic region. Under these assumptions
(\ref{eq:CouplChanEqs2}) reduces to the following set of $N+1$ equations ($%
n=0..N$): 
\begin{align}
\sum_{K^{\prime }}\sum_{m<N}\left\langle n,K\left| \hat{H}-E\right|
m,K^{\prime }\right\rangle c_{m}^{(0)K^{\prime }}-& \sum_{K^{\prime
}}S_{K_{i}K^{\prime }}\left[ \beta _{0}^{(+)K^{\prime }}\ \delta _{n,0}\
\delta _{K^{\prime }K}+V_{n}^{KK^{\prime }\left( +\right) }\right]  \nonumber
\\
& =-\beta _{0}^{(-)K}\ \delta _{n,0}\ \delta _{K_{i}K}-V_{n}^{KK_{i}\left(
-\right) }
\end{align}

The total number of equations for a given entrance channel $K_{i}$ amounts
to $N_{ch}\left( N+1\right) $, and solving the set of equations by
traditional numerical linear algebra leads to the $N_{ch}N$ residual
coefficients $\left\{ c_{n}^{K\left( 0\right) }\text{; }K=K_{\min }..K_{\max
}\text{; }n=0..N-1\right\} $ and $N_{ch}$ $S$-matrix elements $\left\{
S_{K_{i}K}\text{; }K=K_{\min }..K_{\max }\right\} $. The set of equations
has to be solved for all $N_{ch}$ entrance channels labelled by $K_{i}$.

\subsection{Numerical solution and convergence\label{sec:Convergence}}

The numerical solution of the AM equations crucially depends on a proper
choice of $N$, distinguishing the internal from the external region. The
determining factor in this is the form of the potential energy matrix
elements which, contrary to the short-range coordinate character of the
potential, can be of a slowly descending nature in $n$. If this is the case,
a sufficiently large value of $N$ has to be chosen.

In the case of three-cluster systems it is known from literature \cite
{kn:Fedorov94} that the potential asymptotically behaves as $1/\rho ^{3}$ in
the hyperradius, with a corresponding effect on the matrix elements. It will
be shown later on that the asymptotic form of the effective potential in the
current case follows this behavior. It is well known \cite
{kn:Landau,kn:Calogero,kn:Babikov} that potentials with an asymptotic tail $%
1/\rho ^{3}$ dramatically change the phase shift behavior in the low energy
region and that special care should be taken to get convergent results. This
cannot always be obtained by merely choosing a sufficiently large value of $%
N $.

In reference \cite{kn:VA_PR} a numerical strategy to account for long tails
in the potential in the AM was developed. It dramatically improves the
convergence of the results with significantly smaller values for $N$. We
refer to \cite{kn:VA_PR} for further details.

\section{The Generating Function method}

\subsection{General principle}

In this section, the general principles for calculating matrix elements in a
three-cluster basis will be explained. The two main quantities of interest
are: the overlap matrix, and the hamiltonian matrix. The former is of
importance because of the proper normalization of the basis states. The
latter is decomposed into the kinetic energy operator, the matrix elements
of which are obtained mainly by group-theoretical considerations, the
potential energy operator, which in our case will be chosen to be a
semi-realistic two-body interaction based on a superposition of Gaussians,
and the Coulomb contribution.

In this work matrix elements for two-body Gaussian interactions will be
derived. From these matrix elements of other functional forms of two-body
interactions can be obtained using Gaussian transforms. This latter
procedure will be followed to calculate the matrix elements of the Coulomb
interaction.

The basic principle of generating functions is well-known from mathematical
physics. A generating function or generator state depends on a parameter,
referred to as the generating coordinate, in such a way that an expansion
with respect to that parameter yields basis states as expansion terms. A
familiar example are the single-particle translated Gaussian wave functions

\begin{equation}
\phi ({\bf r}\left| {\bf R}\right. )=\exp \left\{ -\frac{1}{2}{\bf r}^{2}+%
\sqrt{2}\ {\bf R}\cdot {\bf r}-\frac{1}{2}{\bf R}^{2}\right\}
\label{eq:sOrbital}
\end{equation}
with the translation parameter ${\bf R}$ acting as generator coordinate. The
choice of parametrization of the generator coordinate influences the quantum
numbers of the individual basis states that are generated. In a Cartesian
parametrization ${\bf R}=(R_{x},R_{y},R_{z})$ one generates the familiar
Cartesian $\phi _{n_{x}}(R_{x})\phi _{n_{y}}(R_{y})\phi _{n_{z}}(R_{z})$
oscillator states. With a radial parametrization ${\bf R}=R{\bf \check{R}}$
(where the inverted hat stands for a unit vector) the expansion yields 
\begin{equation}
\phi ({\bf r}\left| {\bf R}\right. )=\sum_{n,l,m}{\cal N}_{nl}R^{2n+l}Y_{lm}(%
{\bf \check{R})}\phi _{nlm}({\bf r})  \label{eq:sOrbitalRad}
\end{equation}
An underlying mathematical connection exists between such expansions, group
representation theory and coherent state analysis \cite
{perelom72,kn:perelomov}. In the present work we exploit the generating
function principle to facilitate the computation of matrix elements. The
matrix element of any operator between generating states is a function of
the generating coordinates on the left and right 
\begin{equation}
X(\,{\bf R},{\bf R}^{\prime })=\left\langle \phi \left( {\bf r}\left| {\bf R}%
\right. \right) \left| {\bf \hat{X}}\right| \phi \left( {\bf r}\left| {\bf R}%
^{\prime }\right. \right) \right\rangle  \label{eq:GFMEgeneral}
\end{equation}
Expansion of this function will yield the matrix elements between the basis
states. They can be identified in the expansion by the appropriate
dependence on the generator coordinates 
\begin{equation}
X(\,{\bf R},{\bf R}^{\prime })=\sum_{nlm}\sum_{n^{\prime }l^{\prime
}m^{\prime }}{\cal N}_{nl}{\cal N}_{n^{\prime }l^{\prime }}R^{2n+l}R^{\prime
(2n^{\prime }+l^{\prime })}Y_{lm}^{\ast }({\bf \check{R})}Y_{l^{\prime
}m^{\prime }}({\bf \check{R}}^{^{\prime }}{\bf )}\left\langle \phi _{nlm}(%
{\bf r})\left| {\bf \hat{X}}\right| \phi _{n^{\prime }l^{\prime }m^{\prime
}}({\bf r})\right\rangle  \label{eq:GFMEExpand}
\end{equation}
Of course, one is not required to expand with respect to all parameters at
once. Elimination of the angular dependence first, yields a partial
generating function for the radial matrix elements: 
\begin{eqnarray}
X(\,{\bf R},{\bf R}^{\prime }) &=&\sum_{lm}\sum_{l^{\prime }m^{\prime
}}Y_{lm}^{\ast }({\bf \check{R})}Y_{l^{\prime }m^{\prime }}({\bf \check{R}}%
^{^{\prime }}{\bf )}X_{lm;l^{\prime }m^{\prime }}(R,R^{\prime })  \nonumber
\\
X_{lm;l^{\prime }m^{\prime }}(R,R^{\prime }) &=&\sum_{n}\sum_{n^{\prime }}%
{\cal N}_{nl}{\cal N}_{n^{\prime }l^{\prime }}R^{2n+l}R^{\prime (2n^{\prime
}+l^{\prime })}\left\langle \phi _{nlm}({\bf r})\left| {\bf \hat{X}}\right|
\phi _{n^{\prime }l^{\prime }m^{\prime }}({\bf r})\right\rangle
\label{eq:GFRed}
\end{eqnarray}
Such partial generating functions will prove to be particularly useful when
we apply the generating function method to the three-cluster problem with
six generator coordinates (corresponding to six degrees of freedom) and an
extensive set $\nu =\{n,K,(l_{1}l_{2})LM\}$ of quantum numbers.

The calculation of the matrix elements with the generating function method
is a two-step process. The first step is the calculation of the generating
function for the operator involved. Usually this is accomplished with
analytical techniques. The second step is the expansion of the generating
function w.r.t. the generator coordinates. Several approaches have been used
in this respect. Explicit differentiation is one of them. Using recurrence
relations for the expansion terms is another one \cite{kn:AM_AJP}. In any
case, the work involved here is straightforward but extremely tedious; both
approaches are best implemented using algebraic manipulation software such
as Mathematica or Maple.

In this paper we introduce a representation of the generating functions in a
manageable form to obtain explicit matrix elements and their connecting
recurrence relations.

\subsection{Three-cluster generator state}

The customary generator state for the inter-cluster basis functions is given
by (in what follows we shall use small ${\bf q}$ for the Jacobi vectors and
capital ${\bf Q}$ for the corresponding generating coordinates)

\begin{equation}
\Psi \left( {\bf q_{1},q_{2}}\left| {\bf Q_{1},Q_{2}}\right. \right) =\exp
\left\{ -\frac{1}{2}\left( {\bf q}_{1}^{2}+{\bf q_{2}^{2}}\right) +\sqrt{2}%
\left( {\bf Q}_{1}\cdot {\bf q}_{1}+{\bf Q}_{2}\cdot {\bf q}_{2}\right) -%
\frac{1}{2}\left( {\bf Q}_{1}^{2}+{\bf Q}_{2}^{2}\right) \right\}
\label{eq:sorbitalInQ}
\end{equation}
The choice of parametrization is linked to the basis states one intends to
generate. Associated with our choice of basis (Zernike-Brinkman \cite{kn:ZB}%
), we introduce hyperspherical coordinates. The hyperradius and hyperangles,
both for spatial coordinates and for generating parameters, are defined by: 
\begin{eqnarray}
\rho =\sqrt{{\bf q}_{1}^{2}+{\bf q}_{2}^{2}},\, &q_{1}=\rho \cos \theta
,\,&q_{2}=\rho \sin \theta ;  \nonumber \\
R=\sqrt{{\bf Q}_{1}^{2}+{\bf Q}_{2}^{2}},\, &\,Q_{1}=R\cos \Theta
,\,&Q_{2}=R\sin \Theta ,  \label{eq:HyperCoords}
\end{eqnarray}
Using these, one expands the generating function (\ref{eq:sorbitalInQ}) in
hyperspherical harmonic functions: 
\begin{equation}
\Psi \left( {\bf q_{1},q_{2}}\left| {\bf Q_{1},Q_{2}}\right. \right)
=\sum_{\nu }\Psi _{\nu }\left( \rho ,\theta ,{\bf \check{q}}_{1},{\bf \check{%
q}}_{2}\right) \ \Xi _{\nu }^{\ast }\left( R,\Theta ,{\bf \check{Q}}_{1},%
{\bf \check{Q}}_{2}\right)  \label{eq:GFHS}
\end{equation}
where the full set of quantum numbers $\nu $ (introduced previously) is
involved in the summation. The oscillator basis functions are

\begin{equation}
\Psi _{\nu }\left( \rho ,\theta ,{\bf \check{q}}_{1},{\bf \check{q}}%
_{2}\right) ={\cal N}_{n,K\ }\rho ^{K}\exp \{-\rho ^{2}/2\}\
L_{n}^{K+2}(\rho ^{2})\ H_{K}^{(l_{1}l_{2})LM}\left( \theta ,{\bf \check{q}}%
_{1},{\bf \check{q}}_{2}\right)  \label{eq:OscFionHS}
\end{equation}
and the generator coordinate functions are

\begin{equation}
\Xi _{\nu }\left( R,\Theta ,{\bf \check{Q}}_{1},{\bf \check{Q}}_{2}\right) =%
{\cal N}_{n,K}\ R^{K+2n}\ H_{K}^{(l_{1}l_{2})LM}\left( \Theta ,{\bf \check{Q}%
}_{1},{\bf \check{Q}}_{2}\right)  \label{eq:GenCoordHS}
\end{equation}
Here $H$ denotes the hyperspherical harmonic function

\begin{eqnarray}
H_{K}^{(l_{1}l_{2})LM}\left( \Theta ,{\bf \check{Q}}_{1},{\bf \check{Q}}%
_{2}\right) &=&{\cal N}_{K}^{(l_{1}l_{2})LM}\Phi _{K}^{(l_{1}l_{2})}\left(
\Theta \right) \left\{ 
\mathop{\rm Y}%
\nolimits_{l_{1}}({\bf \check{Q}}_{1})\times 
\mathop{\rm Y}%
\nolimits_{l_{2}}({\bf \check{Q}}_{2})\right\} _{LM}  \nonumber \\
\Phi _{K}^{(l_{1}l_{2})}\left( \Theta \right) &=&\left( \cos \Theta \right)
^{l_{1}}\ \left( \sin \Theta \right) ^{l_{2}}\ P_{\frac{K-l_{1}-l_{2}}{2}%
}^{l_{2}+\frac{1}{2},l_{1}+\frac{1}{2}}\,(\cos 2\Theta )  \label{eq:HSFion}
\end{eqnarray}
From (\ref{eq:GenCoordHS}), one easily deduces the procedure for selecting
basis functions with fixed quantum numbers $\nu =\{n,K,\left(
l_{1}l_{2}\right) LM\}$. One has to differentiate the generating function $%
\left( K+2n\right) $-times with respect to $R$ and then to set $R=0$. After
that one has to integrate over $\Theta $ with the weight $\Phi
_{K}^{(l_{1}l_{2})}$ to project onto the hypermomentum $K$; one has to
integrate over unit vectors ${\bf \check{Q}}_{1}$ and ${\bf \check{Q}}_{2}$
with weights $%
\mathop{\rm Y}%
_{l_{1}m_{1}}({\bf \check{Q}}_{1})$ and $%
\mathop{\rm Y}%
_{l_{2}m_{2}}({\bf \check{Q}}_{2})$ to project onto partial angular momenta.
The order of these operations is not important and is a matter of
convenience for each specific case. However these calculations, in
particular those connected with integrating over hyperangle $\Theta $, are
extremely extensive and cumbersome. For this reason, we introduce a new
generating function appropriate for three-cluster calculations.

We start from the function below which depends on seven generating
coordinates namely: 
\begin{eqnarray}
&&\Psi \left( {\bf q_{1},q_{2}}\left| \epsilon ,{\bf Q_{1},Q_{2}}\right.
\right)  \label{eq:sOrbitalInQEps} \\
&=&(1+\epsilon )^{-3}\exp \left\{ -\frac{1}{2}\frac{1-\epsilon }{1+\epsilon }%
\ \left( {\bf q}_{1}^{2}+{\bf q_{2}^{2}}\right) +\frac{\sqrt{2}}{1+\epsilon }%
\left( {\bf Q}_{1}\cdot {\bf q}_{1}+{\bf Q}_{2}\cdot {\bf q}_{2}\right) -%
\frac{1}{2}\frac{1}{1+\epsilon }\left( {\bf Q}_{1}^{2}+{\bf Q}%
_{2}^{2}\right) \right\}
\end{eqnarray}
It was used previously in a different context \cite
{nest_8be,kn:fil_rev3,kn:Vasi92} to describe the coupling between monopole
and two-cluster degrees of freedom. In those cases, the parameter ${\bf Q}$
generates basisfunctions of inter-cluster motion while parameter $\epsilon $
generates collective monopole excitations of the $A$-nucleon system. Here,
we will modify the function somewhat and use it only for the inter-cluster
motion. We exploit the redundancy in the set of generating parameters (seven
parameters vs. six degrees of freedom) and the fact that all expressions up
to now are valid for complex generator coordinates also. We restrict the
moduli of ${\bf Q}_{1}$ and ${\bf Q}_{2}$ and set 
\begin{equation}
Q_{1}=S,\,Q_{2}=-iS  \label{eq:SubstSForQ}
\end{equation}
When the complex conjugate version of (\ref{eq:sOrbitalInQEps}) is used,
e.g. in the calculation of matrix elements, (\ref{eq:SubstSForQ}) is also
complex conjugated.

We now consider the new set of generator coordinates $\epsilon {\bf ,}S{\bf ,%
\check{Q}}_{1}{\bf ,\check{Q}}_{{\bf 2}}$, and substitute (\ref
{eq:SubstSForQ}) in (\ref{eq:sOrbitalInQEps}), where the inverted hats on $%
{\bf Q}_{1}$ and ${\bf Q}_{2}$ again indicate the angular components of both
variables. This leads to 
\begin{eqnarray}
\Psi \left( {\bf q}\left| \epsilon {\bf ,}S{\bf ,\check{Q}}_{1},{\bf \check{Q%
}}_{2}\right. \right) &=&(1+\epsilon )^{-3}\exp \left\{ -\frac{1}{2}\frac{%
1-\epsilon }{1+\epsilon }\ \left( {\bf q}_{1}^{2}+{\bf q_{2}^{2}}\right) +%
\frac{\sqrt{2}S}{1+\epsilon }\left( {\bf \check{Q}}_{1}{\bf \cdot q}_{1}-i%
{\bf \check{Q}}_{2}\cdot {\bf q}_{2}\right) \right\}  \nonumber \\
&=&\sum_{\nu }\Psi _{\nu }\left( \rho ,\theta ,{\bf \check{q}}_{1},{\bf 
\check{q}}_{2}\right) \ \Phi _{\nu }\left( \epsilon ,S,{\bf \check{Q}}_{1}%
{\bf ,\check{Q}}_{2}\right)  \label{eq:sOrbitalsInQEpsS}
\end{eqnarray}
where the weights associated with each basis function are given by 
\begin{equation}
\Phi _{\nu }\left( \epsilon ,S,{\bf \check{Q}}_{1}{\bf ,\check{Q}}%
_{2}\right) ={\cal N}_{K}^{(l_{1}l_{2})LM}{\cal N}_{n,K}\ \left( -i\right)
^{l_{2}}\epsilon ^{n\ }S^{K\ }\left\{ 
\mathop{\rm Y}%
\nolimits_{l_{1}}({\bf \check{Q}}_{1})\times 
\mathop{\rm Y}%
\nolimits_{l_{2}}({\bf \check{Q}}_{2})\right\} _{LM}
\label{eq:sOrbitalInQEpsSWeights}
\end{equation}
These are of a simpler structure and easier to use than (\ref{eq:GenCoordHS}%
) because through \ (\ref{eq:SubstSForQ}) the dependence on the hyperangular
coordinate has been eliminated.

The full generating function for the matrix elements $X_{\nu ,\nu ^{\prime
}} $ of operator ${\bf \hat{X}}$ now has the following general structure: 
\begin{eqnarray}
&&X\left( \epsilon ,S,{\bf \check{Q}}_{1}{\bf ,\check{Q}}_{{\bf 2}};\epsilon
^{\prime },S^{\prime },{\bf \check{Q}}_{1}^{\prime }{\bf ,\check{Q}}_{{\bf 2}%
}^{\prime }\right)  \nonumber \\
&=&\sum_{\nu ,\nu ^{\prime }}\left\langle \Psi _{\nu }\left( \rho ,\theta ,%
{\bf \check{q}}_{1},{\bf \check{q}}_{2}\right) \left| {\bf \hat{X}}\right|
\Psi _{\nu ^{\prime }}\left( \rho ,\theta ,{\bf \check{q}}_{1},{\bf \check{q}%
}_{2}\right) \right\rangle \ \Phi _{\nu }^{\ast }\left( \epsilon ,S,{\bf 
\check{Q}}_{1}{\bf ,\check{Q}}_{2}\right) \Phi _{\nu ^{\prime }}\left(
\epsilon ^{\prime }{\bf ,}S^{\prime }{\bf ,\check{Q}}_{1}^{\prime }{\bf ,%
\check{Q}}_{{\bf 2}}^{\prime }\right)  \nonumber \\
&=&\sum_{\nu ,\nu ^{\prime }}\ X_{\nu ,\nu ^{\prime }}\ {\cal N}%
_{K}^{(l_{1}l_{2})LM}{\cal N}_{K^{\prime }}^{(l_{1}^{\prime }l_{2}^{\prime
})L^{\prime }M^{\prime }}{\cal N}_{n,K}{\cal N}_{n^{\prime },K^{\prime }}\
\left( -i\right) ^{l_{2}}\left( i\right) ^{l_{2}^{\prime }}\epsilon
^{n}\epsilon ^{n^{\prime }}S^{K}S^{\prime K^{\prime }}  \nonumber \\
&&\qquad \qquad \left\{ 
\mathop{\rm Y}%
\nolimits_{l_{1}}({\bf \check{Q}}_{1})\times 
\mathop{\rm Y}%
\nolimits_{l_{2}}({\bf \check{Q}}_{2})\right\} _{LM}\left\{ 
\mathop{\rm Y}%
\nolimits_{l_{1}^{\prime }}({\bf \check{Q}}_{1}^{\prime })\times 
\mathop{\rm Y}%
\nolimits_{l_{2}^{\prime }}({\bf \check{Q}}_{2}^{\prime })\right\}
_{L^{\prime }M^{\prime }}  \label{eq:GFEpsSFull}
\end{eqnarray}
where again the shorthand notation for the quantum numbers $\nu =\left(
n,K,(l_{1}l_{2})LM\right) $ and $\nu ^{\prime }=\left( n^{\prime },K^{\prime
},(l_{1}^{\prime }l_{2}^{\prime })L^{\prime }M^{\prime }\right) $ is used.
As explained before, we will also consider partial generating functions
which have been reduced with respect to a subset of generating coordinates
and their corresponding quantum numbers. Most often, we will use a reduction
with respect to the angular momentum dependence i.e. 
\begin{eqnarray}
&&X\left( \epsilon ,S,{\bf \check{Q}}_{1}{\bf ,\check{Q}}_{{\bf 2}};\epsilon
^{\prime },S^{\prime },{\bf \check{Q}}_{1}^{\prime },{\bf \check{Q}}_{{\bf 2}%
}^{\prime }\right)  \nonumber \\
&=&\sum_{\left( l_{1}l_{2}\right) LM,(l_{1}^{\prime }l_{2}^{\prime
})L^{\prime }M^{\prime }}{\cal X}_{\left( l_{1}l_{2}\right)
LM;(l_{1}^{\prime }l_{2}^{\prime })L^{\prime }M^{\prime }}(\epsilon
,S,\epsilon ^{\prime },S^{\prime })  \nonumber \\
&&\qquad \qquad \left\{ 
\mathop{\rm Y}%
\nolimits_{l_{1}}({\bf \check{Q}}_{1})\times 
\mathop{\rm Y}%
\nolimits_{l_{2}}({\bf \check{Q}}_{2})\right\} _{LM}\left\{ 
\mathop{\rm Y}%
\nolimits_{l_{1}^{\prime }}({\bf \check{Q}}_{1}^{\prime })\times 
\mathop{\rm Y}%
\nolimits_{l_{2}^{\prime }}({\bf \check{Q}}_{2}^{\prime })\right\}
_{L^{\prime }M^{\prime }}  \label{eq:GFEpsSExpandInL}
\end{eqnarray}
with the partial generating function 
\begin{eqnarray}
&&{\cal X}_{\left( l_{1}l_{2}\right) LM;(l_{1}^{\prime }l_{2}^{\prime
})L^{\prime }M^{\prime }}(\epsilon ,S,\epsilon ^{\prime },S^{\prime })= 
\nonumber \\
&&\qquad \sum_{n,K,n^{\prime },K^{\prime }}X_{K,n,\left( l_{1}l_{2}\right)
LM;K,n,(l_{1}^{\prime }l_{2}^{\prime })L^{\prime }M^{\prime }}{\cal N}%
_{K}^{(l_{1}l_{2})LM}{\cal N}_{K^{\prime }}^{(l_{1}^{\prime }l_{2}^{\prime
})L^{\prime }M^{\prime }}{\cal N}_{n,K}{\cal N}_{n^{\prime },K^{\prime
}}\left( -i\right) ^{l_{2}}\left( i\right) ^{l_{2}^{\prime }}\epsilon
^{n}\epsilon ^{\prime n^{\prime }}S^{K}S^{\prime K^{\prime }}
\label{eq:GFEpsSRed}
\end{eqnarray}
generating matrix elements for specified $\left( l_{1}l_{2}\right) LM$ and $%
(l_{1}^{\prime }l_{2}^{\prime })L^{\prime }M^{\prime }$ only.

The asymmetry in the treatment of the different quantum numbers is motivated
by the methodology in which the matrix elements will be used. Indeed, all
quantities in this paper are calculated in the context of the Algebraic
Model. As explained earlier the spatial asymptotic behavior is mapped onto
the asymptotic behavior of the expansion coefficients in the oscillator
basis. As for fixed $K$ the $n$ quantum numbers ladder through the
oscillator shells, they will be needed for sufficiently high values in order
to properly describe the asymptotic region.

\section{Matrix elements in the folding approximation}

In this section we derive the generating functions for the overlap and
hamiltonian in the folding approximation, for several reasons. The folding
approximation is indeed the natural representation for discussing the
asymptotic behavior of the three cluster system, as the antisymmetrization
between clusters vanishes at large inter-cluster distances. The calculation
of generating functions in this approximation is also illuminating for the
subsequent derivation of generating functions in a fully antisymmetrized
setting, as the principles are identical, but the implementation is more
complex. Finally, the folding approximation provides an interesting model to
discuss the importance of antisymmetrization in the interaction region.

\subsection{Matrix elements for the overlap}

The overlap of two generating functions of the form (\ref{eq:sOrbitalInQEps}%
) is easily obtained, and can be written as 
\begin{equation}
I\left( \epsilon ,{\bf Q}_{1},{\bf Q}_{2};\epsilon ^{\prime },{\bf Q}%
_{1}^{\prime },{\bf Q}_{2}^{\prime }\right) \,=\Delta ^{-3}\exp \left\{ 
\frac{1}{\Delta }\sum_{i=1}^{2}\left[ {\bf Q}_{i}\cdot {\bf Q}_{i}^{\prime }+%
\frac{1}{2}\left( \epsilon ^{\prime }{\bf Q}_{i}^{2}+\epsilon {\bf Q}%
_{i}^{\prime 2}\right) \right] \right\}  \label{eq:FoldOvlapEps}
\end{equation}
where 
\[
\Delta =1-\epsilon \epsilon ^{\prime } 
\]

After substitution of (\ref{eq:SubstSForQ}) in (\ref{eq:FoldOvlapEps}), one
obtains 
\begin{equation}
I\left( \epsilon ,S,{\bf Q}_{1},{\bf Q}_{2};\epsilon ^{\prime },S^{\prime },%
{\bf Q}_{1}^{\prime },{\bf Q}_{2}^{\prime }\right) =\Delta ^{-3}\exp \left\{ 
\frac{SS^{\prime }}{\Delta }{\bf \check{Q}}_{1}\cdot {\bf \check{Q}}%
_{1}^{\prime }\right\} \exp \left\{ \frac{SS^{\prime }}{\Delta }{\bf \check{Q%
}}_{2}\cdot {\bf \check{Q}}_{2}^{\prime }\right\}  \label{eq:FoldOvlapEpsS}
\end{equation}

It is interesting to note that the arguments of the exponential factors are
diagonal in the generator coordinates $S$ and $S^{\prime }$. To obtain an
expansion in terms of angular momenta, the well known relation 
\begin{equation}
\exp \{{\bf a\cdot }{\bf b}\}=4\pi \sum_{lm}%
\mathop{\rm i}%
\nolimits_{l}(ab)\,%
\mathop{\rm Y}%
\nolimits_{lm}^{\ast }({\bf \hat{a}})\,%
\mathop{\rm Y}%
\nolimits_{lm}^{\ast }({\bf \hat{b}})  \label{eq:ExpExpansReal}
\end{equation}
can be applied, where $%
\mathop{\rm i}%
_{l}(x)$ is the Modified Spherical Bessel function of the first kind.
Substitution of (\ref{eq:ExpExpansReal}) in (\ref{eq:FoldOvlapEpsS}), and
applying traditional angular momentum coupling techniques leads to 
\begin{eqnarray}
&&I\left( \epsilon ,S,{\bf Q}_{1},{\bf Q}_{2};\epsilon ^{\prime },S^{\prime
},{\bf Q}_{1}^{\prime },{\bf Q}_{2}^{\prime }\right)  \nonumber \\
&=&\left( 4\pi \right) ^{2}\sum_{l_{1},m_{1},l_{2},m_{2}}%
\mathop{\rm i}%
\nolimits_{l_{1}}\left( \frac{SS^{\prime }}{\Delta }\right) 
\mathop{\rm i}%
\nolimits_{l_{2}}\left( \frac{SS^{\prime }}{\Delta }\right) 
\mathop{\rm Y}%
\nolimits_{l_{1}m_{1}}^{\ast }({\bf \check{Q}}_{1})\,%
\mathop{\rm Y}%
\nolimits_{l_{2}m_{2}}^{\ast }({\bf \check{Q}}_{2})%
\mathop{\rm Y}%
\nolimits_{l_{1}m_{1}}^{\ast }({\bf \check{Q}}_{1}^{\prime })%
\mathop{\rm Y}%
\nolimits_{l_{2}m_{2}}^{\ast }({\bf \check{Q}}_{2}^{\prime })  \nonumber \\
&=&\left( 4\pi \right) ^{2}\sum_{\left( l_{1}l_{2}\right) LM}%
\mathop{\rm i}%
\nolimits_{l_{1}}\left( \frac{SS^{\prime }}{\Delta }\right) 
\mathop{\rm i}%
\nolimits_{l_{2}}\left( \frac{SS^{\prime }}{\Delta }\right) \left\{ 
\mathop{\rm Y}%
\nolimits_{l_{1}}({\bf \check{Q}}_{1})\times \,%
\mathop{\rm Y}%
\nolimits_{l_{2}}({\bf \check{Q}}_{2})\right\} _{LM}\left\{ 
\mathop{\rm Y}%
\nolimits_{l_{1}}^{\ast }({\bf \check{Q}}_{1}^{\prime })%
\mathop{\rm Y}%
\nolimits_{l_{2}}^{\ast }({\bf \check{Q}}_{2}^{\prime })\right\} _{LM}
\label{eq:FoldOvlapEpsSExpandLM}
\end{eqnarray}
The reduced generating function (cfr. (\ref{eq:GFEpsSRed})) then becomes 
\begin{equation}
{\cal I}_{\left( l_{1}l_{2}\right) LM;(l_{1}l_{2})LM}\left( \epsilon
,S;\epsilon ^{\prime },S^{\prime }\right) =\left( 4\pi \right) ^{2}\Delta \
^{-3}%
\mathop{\rm i}%
\nolimits_{l_{1}}\left( \frac{SS^{\prime }}{\Delta }\right) 
\mathop{\rm i}%
\nolimits_{l_{2}}\left( \frac{SS^{\prime }}{\Delta }\right)
\label{eq:FoldOvlapRed}
\end{equation}
This reduced generating function (\ref{eq:FoldOvlapRed}) is diagonal in the
partial angular momenta $l_{1}$\ and $l_{2}$, and independent of total
angular momentum $L$, thus valid for all angular momenta $L$ compatible with 
$l_{1}$\ and $l_{2}$.

The matrix elements with quantumnumbers $K$ (also $n$) can now be obtained
through a standard procedure, e.g. by differentiating (\ref{eq:FoldOvlapRed}%
) with respect to $S$ and $S^{\prime }$ ($\epsilon $ and $\epsilon ^{\prime
} $ for $n$). In particular, as (\ref{eq:FoldOvlapRed}) depends on the
product of $S$ and $S^{\prime }$ only, the overlap is diagonal in $K$.
Likewise, the dependence on $\epsilon $ and $\epsilon ^{\prime }$ appears as
a product in the factor $\Delta $, so that the overlap is again diagonal in $%
n$. The fact that the generated overlap matrix elements are diagonal is a
confirmation of the fact that these matrix elements were generated in an
orthogonal basis. The calculation of the matrix is however not unimportant,
as it provides a straightforward way to obtain the norm of the generator
coordinate basisfunction (\ref{eq:sOrbitalInQEpsSWeights}).

\subsection{Matrix elements for the kinetic energy}

In order to calculate the kinetic energy one can use the properties of the
oscillator basis. The matrix elements of the kinetic energy of relative
motion of the clusters are related to those of the oscillator potential by
the virial theorem 
\begin{eqnarray}
\left\langle N\left| \hat{T}_{R}\right| N\right\rangle &=&\left\langle
N\left| \hat{V}_{O}\right| N\right\rangle =\frac{1}{2}E_{N}  \nonumber \\
\left\langle N\pm 2\left| \hat{T}_{R}\right| N\right\rangle &=&-\left\langle
N\pm 2\left| \hat{V}_{O}\right| N\right\rangle
\end{eqnarray}
where 
\begin{equation}
\hat{V}_{O}=\frac{1}{2}\hbar \omega \sum_{i=1}{\bf q}_{i}^{2}
\end{equation}
and $E_{N}=\hbar \omega \left[ N+3\right] =\hbar \omega \left[ 2n+K+3\right] 
$ is the oscillator energy of relative motion.

One easily obtains a representation of the oscillator potential in the
manifold spanned by the generating function (\ref{eq:sOrbitalsInQEpsS}): 
\begin{eqnarray}
&&\left( {\bf q}_{1}^{2}+{\bf q}_{2}^{2}\right) \Psi \left( {\bf q}_{1}{\bf %
,q}_{2}\left| \epsilon ,S;{\bf \check{Q}}_{1}{\bf ,\check{Q}}_{2}\right.
\right)  \nonumber \\
&=&\left[ (1+\epsilon )^{2}\frac{d}{d\epsilon }+(1+\epsilon )S\frac{d}{dS}%
+3(1+\epsilon )\right] \Psi \left( {\bf q}_{1}{\bf ,q}_{2}\left| \epsilon ,S;%
{\bf \check{Q}}_{1}{\bf ,\check{Q}}_{2}\right. \right)
\end{eqnarray}
so that the kinetic energy operator $\hat{T}_{R}$ can be represented by 
\begin{equation}
{\cal T}_{R}=\frac{1}{2}\hbar \omega \left[ -(1-\epsilon )^{2}\frac{d}{%
d\epsilon }+(1-\epsilon )S\frac{d}{dS}+3(1-\epsilon )\right]
\end{equation}

In a wider context, one can associate this operator with the generators $%
{\cal R}^{\left( +\right) }$, ${\cal R}^{\left( -\right) }$ and ${\cal R}%
^{\left( 0\right) }$%
\begin{eqnarray}
{\cal R}^{\left( +\right) } &=&\left[ \epsilon ^{2}\frac{d}{d\epsilon }%
+\epsilon S\frac{d}{dS}+3\epsilon \right] ,  \nonumber \\
{\cal R}^{\left( -\right) } &=&\left[ \frac{d}{d\epsilon }\right]  \nonumber
\\
{\cal R}^{\left( 0\right) } &=&\left[ 2\epsilon \frac{d}{d\epsilon }+S\frac{d%
}{dS}+3\right]
\end{eqnarray}
of the $Sp(2,R)$ group classifying the space of relative motion of
three-particle states, and whose irreducible representations are labelled by
hypermomentum $K$ so that 
\begin{eqnarray}
{\cal V}_{R} &=&\frac{1}{2}\hbar \omega \left[ {\cal R}^{\left( 0\right) }+%
{\cal R}^{\left( +\right) }+{\cal R}^{\left( -\right) }\right]  \nonumber \\
{\cal T}_{R} &=&\frac{1}{2}\hbar \omega \left[ {\cal R}^{\left( 0\right) }-%
{\cal R}^{\left( +\right) }-{\cal R}^{\left( -\right) }\right]
\end{eqnarray}

Using again the shorthand notation $\nu _{0}=(l_{1}l_{2})LM$, matrix
elements are then readily found to be 
\begin{eqnarray}
\left\langle n,K,\nu _{0}\left| \hat{T}_{R}\right| n,K,\nu _{0}\right\rangle
&=&\frac{1}{2}\hbar \omega \left[ 2n+K+3\right]  \nonumber \\
\left\langle n+1,K,\nu _{0}\left| \hat{T}_{R}\right| n,K,\nu
_{0}\right\rangle &=&-\frac{1}{2}\hbar \omega \sqrt{\left( n+1\right) \left(
n+K+3\right) }  \nonumber \\
\left\langle n-1,K,\nu _{0}\left| \hat{T}_{R}\right| n,K,\nu
_{0}\right\rangle &=&-\frac{1}{2}\hbar \omega \sqrt{n\left( n+K+2\right) }
\end{eqnarray}

The full kinetic energy of the three cluster system must include the
internal kinetic energy of the clusters: $\hat{T}=\hat{T}_{R}+\hat{T}_{cl}$.
As we consider frozen $s$-clusters only, this contribution is purely
diagonal and equal to 
\begin{equation}
T_{cl}=\frac{3}{4}\hbar \omega \sum_{i=1}^{3}\left( A_{i}-1\right) =\frac{3}{%
4}\hbar \omega \left( A-3\right)
\end{equation}

\subsection{Matrix elements for a Gaussian potential}

For a Gaussian two-body interaction with strength $V_{0}$ and range $a$%
\begin{equation}
V({\bf r}_{i},{\bf r}_{j})=V_{0}\exp \left\{ -\frac{({\bf r}_{i}-{\bf r}%
_{j})^{2}}{a^{2}}\right\}  \label{eq:GaussPot}
\end{equation}
the folding potential (\ref{eq:FoldPot}) for (0s)-clusters can be calculated
analytically. The result obtained from (\ref{eq:FoldPotTerm}) is again of a
Gaussian form but with a modified strength and interaction length now also
depending on the oscillator parameter $b$%
\begin{equation}
V^{(F)}({\bf R}_{\tau \upsilon })=V_{0}\,z^{3/2}\exp \left\{ -\frac{z}{a^{2}}%
{\bf R}_{\tau \upsilon }^{2}\right\}  \label{eq:GaussFoldPotTerm}
\end{equation}
\[
z=\left( 1+\frac{b^{2}}{a^{2}}[2-\mu _{\tau \upsilon }^{-1}]\right)
^{-1},\qquad \mu _{\tau \upsilon }=\frac{A_{\tau }A_{\upsilon }}{A_{\tau
}+A_{\upsilon }} 
\]

The matrix element of a Gaussian potential between two generating functions
of the form (\ref{eq:sOrbitalInQEps}) is not necessarily diagonal in terms
of the chosen Jacobi coordinates. In the folding approximation one can
however easily find a set Jacobi coordinates in which the potential matrix
element is diagonal. Both sets of Jacobi coordinates will then be related by
an orthogonal transformation.

In what follows, we distinguish the two types of Jacobi coordinates as
follows: the original coordinates are denoted by ${\bf q}$ and were
introduced in section \ref{sec:WaveFions}, and shown in Fig. \ref
{fig:figure1}; the diagonalizing coordinates will be denoted by ${\bf x}$.
An explicit instance of the latter are easily obtained. Indeed, for any
choice of two clusters $i$ and $j$ ($i\neq j)$, $k$ being the third
particle, one obtains a system of coordinates ${\bf x}$, uniquely defined by
index $k$, as follows,

\begin{eqnarray}
{\bf x}_{1} &=&\sqrt{c_{1,k}}({\bf r}_{i}-{\bf r}_{j})  \nonumber \\
{\bf x}_{2} &=&\sqrt{c_{2,k}}({\bf r}_{k}-\frac{A_{i}{\bf r}_{i}+A_{j}{\bf r}%
_{j}}{A_{i}+A_{j}})  \label{eq:xJacobi}
\end{eqnarray}
\begin{equation}
c_{1,k}=\mu _{ij}=\frac{A_{i}\,A_{j}}{A_{i}+A_{j}},\qquad c_{2,k}=\frac{%
A_{k}\,(A_{i}+A_{j})}{A_{i}+A_{j}+A_{k}}
\end{equation}
Each such Jacobi coordinate system $k$ leads to a diagonal representation
for the potential energy between clusters $i$ and $j$. As both Jacobi
systems ${\bf q}$ and ${\bf x}$ are related through an orthogonal
transformation, we can invoke the Raynal-Revai theorem \cite{kn:rayn_rev}.
The latter states that any orthogonal transformation of Jacobi coordinates
leads to an orthogonal transformation of the Hyperspherical Harmonics (\ref
{eq:HSFion}) in the wave function (\ref{eq:OscFionHS}), preserving the
hypermomentum quantum number $K$ 
\begin{equation}
H_{K}^{\left( l_{1}l_{2}\right) LM}(\theta _{{\bf q}},{\bf \check{q}}_{1},%
{\bf \check{q}}_{2})=\sum_{\lambda _{1}\lambda _{2}}O_{\lambda _{1}\lambda
_{2}}^{l_{1}l_{2}}H_{K}^{\left( \lambda _{1}\lambda _{2}\right) LM}(\theta _{%
{\bf x}},{\bf \check{x}}_{1},{\bf \check{x}}_{2})  \label{eq:RRTheorem}
\end{equation}
The $O_{\lambda _{1}\lambda _{2}}^{l_{1}l_{2}}$ are known as the
Raynal-Revai coefficients \cite{kn:rayn_rev}.

This transformation can then be used to obtain the matrix elements of the
potential in the original set of coordinates ${\bf q}$ through 
\begin{eqnarray}
&&\left\langle n,K,\left( l_{1}l_{2}\right) LM\left| \hat{V}\right|
n^{\prime },K^{\prime },\left( l_{1}^{\prime }l_{2}^{\prime }\right)
LM\right\rangle _{({\bf q})}  \label{eq:RRTransfo} \\
&=&\sum_{\lambda _{1}\lambda _{2}}\sum_{\lambda _{1}^{\prime }\lambda
_{2}^{\prime }}O_{\lambda _{1}\lambda _{2}}^{l_{1}l_{2}}O_{\lambda
_{1}^{\prime }\lambda _{2}^{\prime }}^{l_{1}^{\prime }l_{2}^{\prime
}}\left\langle n,K,\left( \lambda _{1}\lambda _{2}\right) LM\left| \hat{V}%
\right| n^{\prime },K^{\prime },\left( \lambda _{1}^{\prime }\lambda
_{2}^{\prime }\right) LM\right\rangle _{({\bf x})}
\end{eqnarray}
where the matrix element on the right hand side, which is diagonal in the $%
{\bf x}$ representation, can be calculated in a straightforward way.

To obtain the latter matrix element we consider a generating function of an
identical structure as (\ref{eq:sOrbitalInQEps}), in which we replace the $%
{\bf q}$ coordinates and ${\bf Q}$ generator coordinates by ${\bf x}$ and $%
{\bf X}$. A generating function for the two-body matrix elements is then
easily obtained as a product of two integrals over ${\bf x}_{1}$ and ${\bf x}%
_{2}$ leading to 
\begin{equation}
V\left( \epsilon {\bf ,}S{\bf ,\check{X}}_{1}{\bf ,\check{X}}_{{\bf 2}%
};\epsilon ^{\prime }{\bf ,}S^{\prime }{\bf ,\check{X}}_{1}^{\prime }{\bf ,%
\check{X}}_{{\bf 2}}^{\prime }\right) =V_{pre}\exp \left\{ \frac{SS^{\prime }%
}{\Lambda }{\bf \check{X}}_{1}\cdot {\bf \check{X}}_{1}^{\prime }\right\}
\exp \left\{ \frac{SS^{\prime }}{\Delta }{\bf \check{X}}_{{\bf 2}}\cdot {\bf 
\check{X}}_{{\bf 2}}^{\prime }\right\}  \label{eq:FoldGaussEpsS}
\end{equation}
\begin{eqnarray*}
V_{pre} &=&V_{0}\,\left( \Delta \Lambda \right) ^{-3/2}\exp \left\{ -\frac{%
\gamma }{2\Delta \Lambda }\left[ \left( \xi ^{\prime }S\right) ^{2}+\left(
\xi S^{\prime }\right) ^{2}\right] \right\} \\
\xi &=&1+\epsilon ,\qquad \xi ^{\prime }=1+\epsilon ^{\prime } \\
\Delta &=&1-\epsilon \epsilon ^{\prime }=\xi +\xi ^{\prime }-\xi \xi
^{\prime } \\
\Lambda &=&\Delta +\gamma \xi \xi ^{\prime } \\
\gamma &=&\frac{b^{2}}{a^{2}}\ \frac{1}{c_{1,\alpha }}
\end{eqnarray*}

\smallskip By using (\ref{eq:ExpExpansReal}), one again eliminates
factorized terms and sums to obtain the following reduced generating
function (cfr. (\ref{eq:GFEpsSRed})), in complete analogy to the procedure
taken for (\ref{eq:FoldOvlapRed}): 
\begin{equation}
{\cal V}_{\left( l_{1}l_{2}\right) LM;(l_{1}l_{2})LM}\left( \epsilon
,S;\epsilon ^{\prime },S^{\prime }\right) =V_{pre}%
\mathop{\rm i}%
\nolimits_{l_{1}}\left( \frac{SS^{\prime }}{\Lambda }\right) 
\mathop{\rm i}%
\nolimits_{l_{2}}\left( \frac{SS^{\prime }}{\Delta }\right)
\label{eq:FoldGaussRed}
\end{equation}
As was the case with the overlap, the matrix elements of the potential in
the ${\bf x}$ coordinate system are diagonal with respect to the partial
angular momenta $l_{1}$ and $l_{2}$ and do not depend directly upon the
total angular momentum $L$ ; this is a direct consequence of the
characteristics of the operator. Expression (\ref{eq:FoldGaussRed}) is again
valid for all values of the total angular momenta $L$ that are compatible
with the given partial angular momenta $l_{1}$ and $l_{2}$, and thus
represents a generating function for matrix elements with specific total
angular momentum $L$.

One obtains matrix elements with specific $K$ and $n$ quantumnumbers through
the standard procedures (differentiation, recurrence relations, ...). So for
example one obtains 
\begin{eqnarray}
&&\left\langle n,K,\left( l_{1}l_{2}\right) LM\left| \hat{V}\right|
n^{\prime },K^{\prime },\left( l_{1}l_{2}\right) LM\right\rangle _{({\bf x})}
\label{eq:MatrElemVbyDiffer} \\
&=&\left. \left( \frac{d}{d\epsilon }\right) ^{n}\left( \frac{d}{d\epsilon
^{\prime }}\right) ^{n^{\prime }}\left( \frac{d}{dS}\right) ^{K}\left( \frac{%
d}{dS^{\prime }}\right) ^{K^{\prime }}{\cal V}_{\left( l_{1}l_{2}\right)
LM;(l_{1}l_{2})LM}\left( \epsilon ,S;\epsilon ^{\prime },S^{\prime }\right)
\right| \Sb S=S^{\prime }=0  \\ \epsilon =\epsilon ^{\prime }=0  \endSb 
\end{eqnarray}

\subsection{Matrix elements for the Coulomb potential}

The Coulomb interaction in the folding approximation between two clusters
with $Z_{\tau }$ and $Z_{\upsilon }$ number of protons is given by (\ref
{eq:FoldPotTerm}) as 
\begin{equation}
V_{C}^{(F)}({\bf R}_{\tau \upsilon })=Z_{\tau }Z_{\upsilon }e^{2}\sum_{i\in
A_{\tau }}\sum_{j\in A_{\upsilon }}\int d\tau _{\tau }d\tau _{\upsilon
}|\Psi _{\tau }(A_{\tau })|^{2\ }\frac{1}{\left| {\bf r}_{i}-{\bf r}_{j}+%
{\bf R}_{\tau \upsilon }\right| }|\ \Psi _{\upsilon }(A_{\upsilon })|^{2}
\label{eq:CoulombInFoldApp}
\end{equation}

A straightforward calculation of its matrix elements is impractical and very
tedious. By however using the following Gauss-transform 
\begin{equation}
\frac{1}{r}=\frac{2}{\sqrt{\pi }}\int_{0}^{\infty }dx\exp \{-r^{2}x^{2}\}
\end{equation}
one rewrites (\ref{eq:CoulombInFoldApp}) as 
\begin{eqnarray}
V_{C}^{(F)}(\tau \upsilon ) &=&\frac{2Z_{\tau }Z_{\upsilon }e^{2}}{\sqrt{\pi 
}}\sum_{i\in A_{\tau }}\sum_{j\in A_{\upsilon }}\int d\tau _{\tau }d\tau
_{\upsilon }|\Psi _{\tau }(A_{\tau })|^{2\ }\int_{0}^{\infty }dx\exp
\{-\left( {\bf r}_{i}-{\bf r}_{j}+{\bf R}_{\tau \upsilon }\right)
^{2}x^{2}\}|\ \Psi _{\upsilon }(A_{\upsilon })|^{2}  \nonumber \\
&=&\frac{2Z_{\tau }Z_{\upsilon }e^{2}}{\sqrt{\pi }b}\int_{0}^{\infty
}d\gamma \,z^{3/2}\exp \left\{ -\frac{R_{\tau \upsilon }^{2}}{b^{2}}\gamma
^{2}z\right\}  \label{eq:CoulombInFoldApprInfInt}
\end{eqnarray}
where 
\begin{equation}
z=\left( 1+t\gamma ^{2}\right) ^{-1},\quad t=2-\mu _{\tau \upsilon
}^{-1},\quad \gamma =bx
\end{equation}
and its matrix elements can be obtained by integrating matrix elements
(depending on $z$) of the Gaussian potential.

Introducing the integration variable $s$%
\begin{equation}
s=\frac{t\gamma ^{2}}{1+t\gamma ^{2}},
\end{equation}
transforms (\ref{eq:CoulombInFoldApprInfInt}) to 
\begin{equation}
V_{C}^{(F)}(\tau \upsilon )=\frac{2Z_{\tau }Z_{\upsilon }e^{2}}{\sqrt{\pi }b}%
\frac{1}{2\sqrt{t}}\int_{0}^{1}ds\,s^{-\frac{1}{2}}\,\exp \left\{ -\frac{%
R_{\tau \upsilon }^{2}}{t\,b^{2}}s\right\} =\frac{2Z_{\tau }Z_{\upsilon
}e^{2}}{\sqrt{\pi }b}%
\mathop{\rm erf}%
(\frac{R_{\tau \upsilon }^{2}}{t\,b^{2}})  \label{eq:CoulombInFoldAppErfForm}
\end{equation}
This form shows that the integration can be reduced to a finite interval. It
also shows that the Coulomb interaction between clusters does not behave as $%
R_{\tau \upsilon }^{-1}$ as could be expected. For very large value of $%
R_{\tau \upsilon }$ however (\ref{eq:CoulombInFoldAppErfForm}) properly
reduces to $\frac{Z_{\tau }Z_{\upsilon }e^{2}}{R_{\tau \upsilon }}$, due to
the asymptotic form of the error function.

\subsection{Asymptotic behavior of the potential contributions}

As the folding model is used for defining the asymptotic channels, it is
clear that the asymptotic behavior of the potential energy matrix elements
in this model will be of vital importance for the rate of convergence of the
AM solutions. The effective potential in terms of the hyperradius $\rho $\
is defined by integrating the folding potential over all hyperangles as
formally indicated by 
\begin{eqnarray}
W(\rho ) &=&\sum_{\tau ,\upsilon }W_{\tau \upsilon }(\rho )\equiv
W_{K,l_{1},l_{2}}(\rho )  \nonumber \\
&=&\left\langle K,l_{1},l_{2}\left| \sum_{\tau ,\upsilon }\hat{V}({\bf R}%
_{\tau \upsilon })\right| K,l_{1},l_{2}\right\rangle
\end{eqnarray}
Its asymptotic behavior is then obtained for large values of $\rho $.

In the diagonal representation with Jacobi coordinates ${\bf x}$ the
calculation of $W(\rho )$ between one pair of clusters $\tau $ and $\upsilon 
$ amounts to a straightforward integration over the hyperangles.

The Gaussian interaction then exhibits the following asymptotic behavior 
\begin{equation}
W_{\tau \upsilon }(\rho )\approx
(-1)^{K-l_{1}-l_{2}}V_{0}N_{K}^{l_{1}l_{2}}N_{K}^{l_{1}l_{2}}\left( \frac{1}{%
2}\right) ^{K+3}\Gamma (l_{1}+\frac{3}{2})\left( \frac{\sqrt{\mu _{\tau
\upsilon }}\,a}{\rho }\right) ^{2l_{1}+3}\left( 
\begin{array}{c}
\frac{K+l_{1}-l_{2}+1}{2} \\ 
K-l_{1}-l_{2}
\end{array}
\right) \left( 
\begin{array}{c}
\frac{K+l_{1}-l_{2}+1}{2} \\ 
K-l_{1}-l_{2}
\end{array}
\right)
\end{equation}
which indeed shows a worst-case behavior (for $l_{1}=0$) of the form $1/\rho
^{3}$ as predicted in section \ref{sec:Convergence}.

The analogous evaluation for the Coulomb interaction leads to the following
(exact) expression 
\begin{equation}
W_{\tau \upsilon }(\rho )=\frac{Z_{K,l_{1},l_{2}}^{eff}}{\rho }
\end{equation}
with 
\begin{eqnarray}
Z_{K,l_{1},l_{2}}^{eff} &=&\frac{1}{2}Z_{\tau }Z_{\upsilon }e^{2}\sqrt{\mu
_{\tau \upsilon }}  \nonumber \\
&&\sum_{n,m=0}^{\frac{K-l_{1}-l_{2}}{2}}(-)^{K-l_{1}-l_{2}-n-m}\left( 
\begin{array}{c}
\frac{K-l_{1}+l_{2}+1}{2} \\ 
n
\end{array}
\right) \left( 
\begin{array}{c}
\frac{K-l_{1}+l_{2}+1}{2} \\ 
m
\end{array}
\right) \left( 
\begin{array}{c}
\frac{K+l_{1}-l_{2}+1}{2} \\ 
\frac{K-l_{1}-l_{2}}{2}-n
\end{array}
\right) \left( 
\begin{array}{c}
\frac{K+l_{1}-l_{2}+1}{2} \\ 
\frac{K-l_{1}-l_{2}}{2}-m
\end{array}
\right)  \nonumber \\
&&B(K-l_{1}-n-m+\frac{3}{2},n+m+l_{1}+1)
\end{eqnarray}
and $B$ stands for the beta-function.

These results corroborate the fact that special care should be taken to get
properly convergent results, even more critically when a Coulomb
contribution between the clusters is present.

\section{Matrix elements with Full Antisymmetrization}

When considering full antisymmetrization between all particles of the
three-cluster wave function the normalization of the basis states becomes a
non-trivial problem. The overlap, i.e. the matrix representation of the
antisymmetrization operator ${\cal A}$, needs to be explicitly calculated.

As was carried out in the previous section one would normally start from the
generating state (\ref{eq:sOrbitalInQEps}) to obtain generating matrix
elements in terms of $\epsilon $ and $\epsilon ^{\prime }$, facilitating the
treatment of the hyperangular coordinates. We propose an alternative
representation for the scaled generating state (\ref{eq:sOrbitalInQEps})
which is more suited to our calculations, and hereto introduce the following
integral transformation 
\begin{equation}
\Psi \left( {\bf q}\left| \epsilon ,\,{\bf Q}\right. \right) \,=\,\int \,d%
{\bf k}\,\exp \{-k^{2}\}\,\ \Psi \left( {\bf q;Q}+\sqrt{2\epsilon }{\bf k}%
\right)  \label{eq:EpsScaling}
\end{equation}
This allows to obtain (\ref{eq:sOrbitalInQEps}) by scaling on the generator
coordinate only. In other words, generating matrix elements can be obtained
with the simpler generating state (\ref{eq:sOrbital}) and scaled later on,
reducing effectively the calculational burden.

\subsection{Matrix elements for the overlap}

Because the individual cluster states are Slater determinants one can use
the familiar determinantal formulae to calculate the generating function.
Starting then from the generating state (\ref{eq:sOrbital}) for the
single-particle orbitals one obtains 
\begin{equation}
I\left( {\bf Q}_{1},{\bf Q}_{2};{\bf Q}_{1}^{\prime },{\bf Q}_{2}^{\prime
}\right) \,=\,\sum_{\nu }D^{(\nu )}\exp \left\{ \sum_{i,j=1}^{2}B_{ij}^{(\nu
)}\ {\bf Q}_{i}\cdot {\bf Q}_{j}^{\prime }\right\}
\label{eq:AntisymOvlapInQ}
\end{equation}
The coefficients $D^{(\nu )}$ and $B_{ij}^{(\nu )}$ as well as the number of
the terms depend on the specific type of three-cluster configuration, viz.
the number of nucleons per cluster and their spin-isospin quantum numbers.

In order to reduce (\ref{eq:AntisymOvlapInQ}) with respect to the angular
quantum numbers it is profitable to diagonalize the forms in the
exponentials of (\ref{eq:AntisymOvlapInQ}). This can easily be achieved by
diagonalizing its 2 by 2 coefficient matrix $B_{ij}^{(\nu )}$. This again
amounts to making an orthogonal transformation from the original Jacobi
coordinates ${\bf q}$ to new Jacobi coordinates ${\bf x}$. This orthogonal
transformation will induce a corresponding transformation of the basis
functions which can be handled by the Raynal-Revai theorem as discussed in
the previous section. The block-diagonal form of an exponential term can be
written as 
\begin{equation}
\exp \left\{ \sum_{i=1}^{2}\lambda _{i}\ {\bf X}_{i}\cdot {\bf X}%
_{i}^{\prime }\right\}  \label{eq:AntisymOvlapInX}
\end{equation}

We now introduce the scaling on $\epsilon $ and $\epsilon ^{\prime }$ by
carrying out the transformation (\ref{eq:EpsScaling}) on both generator
coordinates ${\bf X}$ and ${\bf X}^{\prime }$ for every (block-diagonal)
term in (\ref{eq:AntisymOvlapInQ}), leading to 
\begin{equation}
I\left( \epsilon ,{\bf X}_{1},{\bf X}_{2};\epsilon ^{\prime },{\bf X}%
_{1}^{\prime },{\bf X}_{2}^{\prime }\right) =\left( \Delta _{1}\Delta
_{2}\right) ^{-3/2}\exp \left\{ \sum_{i=1}^{2}\frac{\lambda _{i}}{\Delta _{i}%
}\left[ {\bf X}_{i}\cdot {\bf X}_{i}^{\prime }+\frac{\lambda _{i}}{2}\left(
\epsilon ^{\prime }{\bf X}_{i}^{2}+\epsilon {\bf X}_{i}^{\prime 2}\right) %
\right] \right\}  \label{eq:AntisymOvlapInEps}
\end{equation}
where 
\[
\Delta _{i}=1-\lambda _{i}^{2}\epsilon \epsilon ^{\prime } 
\]
In order to further reduce the generating function in the partial angular
momenta $l_{1}$ and $l_{2}$ we substitute (\ref{eq:SubstSForQ}) in (\ref
{eq:AntisymOvlapInEps}) 
\begin{eqnarray}
&&I\left( \epsilon ,S,{\bf \check{X}}_{1},{\bf \check{X}}_{2};\epsilon
^{\prime },S^{\prime },{\bf \check{X}}_{1}^{\prime },{\bf \check{X}}%
_{2}^{\prime }\right)  \nonumber \\
&=&\left( \Delta _{1}\Delta _{2}\right) ^{-3/2}\exp \left\{ \frac{\lambda
_{1}^{2}-\lambda _{2}^{2}}{2\Delta _{1}\Delta _{2}}\left[ S^{2}\epsilon
^{\prime }+S^{\prime 2}\epsilon \right] \right\} \exp \left\{ \frac{\lambda
_{1}}{\Delta _{1}}SS^{\prime }{\bf \check{X}}_{1}\cdot {\bf \check{X}}%
_{1}^{\prime }\right\} \exp \left\{ \frac{\lambda _{2}}{\Delta _{2}}%
SS^{\prime }{\bf \check{X}}_{2}\cdot {\bf \check{X}}_{2}^{\prime }\right\}
\end{eqnarray}

By using (\ref{eq:ExpExpansReal}) this produces in complete analogy to (\ref
{eq:FoldOvlapRed})\ a generating function for all total angular momenta $L$
compatible with $l_{1}$ and $l_{2}$: 
\begin{eqnarray}
&&{\cal I}_{\left( l_{1},l_{2}\right) LM;(l_{1}l_{2})LM}\left( \epsilon
,S;\epsilon ^{\prime },S^{\prime }\right)  \nonumber \\
&=&\left( 4\pi \right) ^{2}\left( \Delta _{1}\Delta _{2}\right) ^{-3/2}\exp
\left\{ \frac{\lambda _{1}^{2}-\lambda _{2}^{2}}{2\Delta _{1}\Delta _{2}}%
\left[ S^{2}\epsilon ^{\prime }+S^{\prime 2}\epsilon \right] \right\} 
\mathop{\rm i}%
\nolimits_{l_{1}}\left( \frac{\lambda _{1}}{\Delta _{1}}SS^{\prime }\right) 
\mathop{\rm i}%
\nolimits_{l_{2}}\left( \frac{\lambda _{2}}{\Delta _{2}}SS^{\prime }\right)
\label{eq:AntisymOvlapRed}
\end{eqnarray}

If $\lambda _{1}=\lambda _{2}$, the overlap depends on the factor $%
SS^{\prime }$ only and is therefore diagonal with respect to hypermomentum $%
K $, as was the case in the folding approximation. In the general case one
will have $\lambda _{1}\neq \lambda _{2}$, and (\ref{eq:AntisymOvlapRed})
generates nondiagonal matrix elements in $K$. Hypermomentum is thus no
longer a good quantum number for three-cluster systems, contrary to the
folding approximation.

The generating function for the overlap with $K=0$, and consequently for
angular momenta $l_{1}=l_{2}=L=0$, is obtained immediately by putting $%
S=S^{\prime }=0$. This simplicity is an indication of the very suitable form
of our generating function.

The matrix elements $I_{n,n^{\prime }}^{K,\nu _{0};K^{\prime },\nu _{0}}$
can now again be generated by a standard procedure, such as differentiation
or recurrence relations.

\subsection{Matrix elements for the kinetic energy}

The matrix elements of the kinetic energy operator can be derived without
use of a generating function. One of the effects of the antisymmetrization
operator is to mix basis states within a fixed oscillator shell. The
``diagonal'' (i.e. within a shell) and ``off-diagonal'' (within neighboring
shells) kinetic energy matrix elements are easily found to be connected to
the matrix elements of the overlap by 
\[
T_{n,n^{\prime }}^{K,\nu _{0};K^{\prime },\nu _{0}}=\frac{1}{2}\hbar \omega %
\left[ 2n+K+3+\frac{3}{2}\left( A-3\right) \right] I_{n,n^{\prime }}^{K,\nu
_{0};K^{\prime },\nu _{0}}\delta _{2n+K,2n^{\prime }+K^{\prime }} 
\]
\[
T_{n+1,n^{\prime }}^{K,\nu _{0};K^{\prime },\nu _{0}}=-\frac{1}{2}\hbar
\omega \sqrt{\left( n+1\right) \left( n+K+3\right) }I_{n,n^{\prime }}^{K,\nu
_{0};K^{\prime },\nu _{0}}\delta _{2n+K,2n^{\prime }+K^{\prime }} 
\]
where the restriction on the quantum numbers to remain on the same
oscillator shell ($2n+K=2n^{\prime }+K^{\prime }$) has been accounted for.

\subsection{Matrix elements for a Gaussian potential}

If one considers a Gaussian form for the nucleon-nucleon potential, and
calculates a generating matrix element for the interaction using Slater
determinants with individual orbitals of the form (\ref{eq:sOrbital}) one
obtains terms of the form 
\begin{equation}
V\left( {\bf Q}_{1}{\bf ,Q}_{{\bf 2}};{\bf Q}_{1}^{\prime }{\bf ,Q}_{{\bf 2}%
}^{\prime }\right) =V_{0}\,\left( 1-\zeta \right) ^{3/2}\exp \left\{ -\zeta %
\left[ \sum_{i=1}^{2}\left( C_{i}{\bf Q}_{i}+C_{i}^{\prime }{\bf Q}%
_{i}^{\prime }\right) \right] ^{2}+\sum_{i,j=1}^{2}B_{ij}\ {\bf Q}_{i}\cdot 
{\bf Q}_{j}^{\prime }\right\}  \label{eq:AntisymGaussInQ}
\end{equation}
where 
\begin{equation}
\zeta =\frac{2b^{2}}{2b^{2}+a^{2}}
\end{equation}
and $V_{0}$ stands for any of the even ($V_{13}$ and $V_{31}$) and odd ($%
V_{11}$ and $V_{33}$) components of the $NN$--interaction. Again $b$ is the
oscillator radius and $a$ the range of the potential well. The first term in
the exponent of (\ref{eq:AntisymGaussInQ}) contains the factor $\zeta $ and
the vectors $C_{1}{\bf Q}_{1}+C_{2}{\bf Q}_{2}$ and $C_{1}^{\prime }{\bf Q}%
_{1}^{\prime }+C_{2}^{\prime }{\bf Q}_{2}^{\prime }$. The latter have a
simple meaning: they define the distance between of the two clusters which
respectively contain one of the nucleons in the interacting pair.

In the diagonal representation (\ref{eq:AntisymGaussInQ}) becomes 
\begin{equation}
V\left( {\bf X}_{1},{\bf X}_{2};{\bf X}_{1}^{\prime }{\bf ,X}_{{\bf 2}%
}^{\prime }\right) =V_{0}\,\exp \left\{ -\zeta \left[ \sum_{i=1}^{2}\left(
G_{i}{\bf X}_{i}+G_{i}^{\prime }{\bf X}_{i}^{\prime }\right) \right]
^{2}+\sum_{i=1}^{2}\lambda _{i}\ {\bf X}_{i}\cdot {\bf X}_{i}^{\prime
})\right\}  \label{eq:AntisymGaussInX}
\end{equation}
and the coefficients $C_{i}$, $C_{i}^{\prime }$ and $G_{i}$, $G_{i}^{\prime
} $ are trivially related by the orthogonal diagonalizing transformation.

In order to eliminate crossterms in ${\bf X}_{1}\cdot {\bf X}_{2}$ and ${\bf %
X}_{1}^{\prime }\cdot {\bf X}_{2}^{\prime }$ we introduce an additional
transformation through the integral identity: 
\begin{equation}
e^{-{\bf a}^{2}}=\frac{1}{\pi ^{3/2}}\int_{-\infty }^{\infty }d{\bf Z}\exp
\left\{ -Z^{2}+2i({\bf Z\bullet }{\bf a})\right\}
\label{eq:IntegralTransform}
\end{equation}
leading to the following integral form for the block (\ref
{eq:AntisymGaussInX}) 
\begin{eqnarray}
&&V\left( {\bf X}_{1},{\bf X}_{2};{\bf X}_{1}^{\prime }{\bf ,X}_{{\bf 2}%
}^{\prime }\right)  \nonumber \\
&=&V_{0}\frac{1}{\pi ^{3/2}}\exp \left\{ \sum_{i=1}^{2}\lambda _{i}\ {\bf X}%
_{i}\cdot {\bf X}_{i}^{\prime }\right\} \int_{-\infty }^{\infty }d{\bf Z\ }%
\exp \left\{ -Z^{2}+2i\zeta \sum_{i=1}^{2}\left( G_{i}\ {\bf Z\cdot X}%
_{i}+G_{i}^{\prime }\ {\bf Z\cdot X}_{i}^{\prime }\right) \right\}
\label{eq:AntisymGaussInXIntegr}
\end{eqnarray}

Using (\ref{eq:EpsScaling}) on both ${\bf X}_{i}$ and ${\bf X}_{i}^{\prime }$%
, one obtains after integrating over the corresponding ${\bf k}_{i}$ and $%
{\bf k}_{i}^{\prime }$ : 
\begin{eqnarray}
&&V\left( \epsilon ;{\bf X}_{1},{\bf X};\epsilon ^{\prime };{\bf X}%
_{1}^{\prime }{\bf ,X}_{{\bf 2}}^{\prime }\right) =\left( \Delta _{1}\Delta
_{2}\right) ^{-3/2}\exp \left\{ \sum_{i=1}^{2}\frac{\lambda _{i}}{2\Delta
_{i}}\left( \epsilon ^{\prime }\ \lambda _{i}{\bf X}_{i}^{2}+\epsilon \
\lambda _{i}{\bf X}_{i}^{\prime 2}+2\ {\bf X}_{i}\cdot {\bf X}_{i}^{\prime
}\right) \right\}  \nonumber \\
&&\frac{1}{\pi ^{3/2}}\int_{-\infty }^{\infty }d{\bf Z}\exp \left\{ -\Delta 
{\bf Z}^{2}+2i\zeta \sum_{i=1}^{2}\left( \xi _{i}\ {\bf Z\cdot X}_{i}+\xi
_{i}^{\prime }\ {\bf Z\cdot X}_{i}^{\prime }\right) \right\}
\label{eq:AntisymGaussEps}
\end{eqnarray}
where again $\Delta _{i}=1-\lambda _{i}^{2}\epsilon \epsilon ^{\prime }$ and 
\begin{eqnarray*}
\xi _{i} &=&\frac{G_{i}+\epsilon ^{\prime }G_{i}^{\prime }\lambda _{i}}{%
\Delta _{i}},\qquad \xi _{i}^{\prime }=\frac{G_{i}^{\prime }+\epsilon
G_{i}\lambda _{i}}{\Delta _{i}} \\
\Delta &=&\left\{ 1+\sum_{i=1}^{2}\frac{1}{\Delta _{i}}\left[ \epsilon
G_{i}^{2}+\epsilon ^{\prime }G_{i}^{\prime 2}+2\epsilon \epsilon ^{\prime
}\lambda _{i}G_{i}G_{i}^{\prime }\right] \right\}
\end{eqnarray*}
To obtain a reduced generating function for specific angular momentum
quantum numbers, we again use expansion formulae of an exponential in terms
of spherical harmonics. For exponential terms with a real scalar product we
use (\ref{eq:ExpExpansReal}), whereas for exponential terms with an
imaginary scalar product we use \cite{kn:varshalov} 
\begin{equation}
\exp \{i\ {\bf a\cdot }{\bf b}\}=4\pi \sum_{lm}i^{l}%
\mathop{\rm j}%
\nolimits_{l}(ab)\,%
\mathop{\rm Y}%
\nolimits_{lm}^{\ast }({\bf \check{a}})\,%
\mathop{\rm Y}%
\nolimits_{lm}^{\ast }({\bf \check{b}})  \label{eq:ExpExpansImag}
\end{equation}
where $%
\mathop{\rm j}%
\nolimits_{l}(ab)$ is the well-known spherical Bessel function.

Applying both (\ref{eq:ExpExpansReal}) and (\ref{eq:ExpExpansImag}) to (\ref
{eq:AntisymGaussEps}) leads to 
\begin{eqnarray}
&&V\left( \epsilon ,{\bf X}_{1},{\bf X};\epsilon ^{\prime },{\bf X}%
_{1}^{\prime }{\bf ,X}_{{\bf 2}}^{\prime }\right) =(4\pi )^{6}\left( \Delta
_{1}\Delta _{2}\right) ^{-3/2}\exp \sum_{i=1}^{2}\left\{ \frac{\lambda
_{i}^{2}}{2\Delta _{i}}\left( \epsilon ^{\prime }\ {\bf X}_{i}^{2}+\epsilon
\ {\bf X}_{i}^{\prime 2}\right) \right\}  \nonumber \\
&&\sum_{k_{1},k_{2},k_{1}^{\prime },k_{2}^{\prime },k_{1}^{\prime \prime
},k_{2}^{\prime \prime }}i^{k_{1}^{\prime }+k_{2}^{\prime }+k_{1}^{\prime
\prime }+k_{2}^{\prime \prime }}\prod_{i=1}^{2}%
\mathop{\rm i}%
\nolimits_{k_{i}}\left( \frac{\lambda _{i}}{\Delta _{i}}X_{i}X_{i}^{\prime
}\right)  \nonumber \\
&&\frac{1}{\pi ^{3/2}}\int_{-\infty }^{\infty }d{\bf Z}\exp \{-\Delta
Z^{2}\}P_{k_{1},k_{2},k_{1}^{\prime },k_{2}^{\prime },k_{1}^{\prime \prime
},k_{2}^{\prime \prime }}\prod_{i=1}^{2}%
\mathop{\rm j}%
\nolimits_{k_{i}^{\prime }}\left( 2\xi _{i}X_{i}Z\right) 
\mathop{\rm j}%
\nolimits_{k_{i}^{\prime \prime }}\left( 2\xi _{i}^{\prime }X_{i}^{\prime
}Z\right)  \label{eq:AntisymGaussEps2}
\end{eqnarray}
Before performing the integration over ${\bf Z}$, we again make the
substitution (\ref{eq:SubstSForQ}) in (\ref{eq:AntisymGaussEps2}) to obtain

\begin{eqnarray}
&&V\left( \epsilon ,S,{\bf \check{X}}_{1},{\bf \check{X}};\epsilon ^{\prime
},S^{\prime },{\bf \check{X}}_{1}^{\prime }{\bf ,\check{X}}_{{\bf 2}%
}^{\prime }\right)  \nonumber \\
&=&(4\pi )^{6}\left( \Delta _{1}\Delta _{2}\right) ^{-3/2}\exp \left\{ \frac{%
\lambda _{1}^{2}-\lambda _{2}^{2}}{2\Delta _{1}\Delta _{2}}\left( \epsilon
^{\prime }S^{2}+\epsilon S^{\prime 2}\right) \right\}  \nonumber \\
&&\sum_{k_{1},k_{2},k_{1}^{\prime },k_{2}^{\prime },k_{1}^{\prime \prime
},k_{2}^{\prime \prime }}i^{k_{1}^{\prime }+k_{2}^{\prime }+k_{1}^{\prime
\prime }+k_{2}^{\prime \prime }}%
\mathop{\rm i}%
\nolimits_{k_{1}}\left( \frac{\lambda _{1}}{\Delta _{1}}SS^{\prime }\right) 
\mathop{\rm i}%
\nolimits_{k_{2}}\left( \frac{\lambda _{2}}{\Delta _{2}}SS^{\prime }\right) 
\nonumber \\
&&\frac{1}{\pi ^{3/2}}\int_{-\infty }^{\infty }d{\bf Z}\exp \{-\Delta Z^{2}\}%
\mathop{\rm j}%
\nolimits_{k_{1}^{\prime }}\left( 2\xi _{1}SZ\right) 
\mathop{\rm j}%
\nolimits_{k_{1}^{\prime \prime }}\left( 2\xi _{1}^{\prime }S^{\prime
}Z\right) 
\mathop{\rm j}%
\nolimits_{k_{2}^{\prime }}\left( -i2\xi _{2}SZ\right) 
\mathop{\rm j}%
\nolimits_{k_{2}^{\prime \prime }}\left( i2\xi _{2}^{\prime }S^{\prime
}Z\right) P_{k_{1},k_{2},k_{1}^{\prime },k_{2}^{\prime },k_{1}^{\prime
\prime },k_{2}^{\prime \prime }}  \label{eq:PotGenFunIntZfinal}
\end{eqnarray}
where: 
\begin{eqnarray}
P_{k_{1},k_{2},k_{1}^{\prime },k_{2}^{\prime },k_{1}^{\prime \prime
},k_{2}^{\prime \prime }} &=&\prod_{i=1}^{2}\left( 
\mathop{\rm Y}%
\nolimits_{k_{i}}({\bf \check{X}}_{i})\cdot 
\mathop{\rm Y}%
\nolimits_{k_{i}}({\bf \check{X}}_{i}^{\prime })\right) \left( 
\mathop{\rm Y}%
\nolimits_{k_{i}^{\prime }}({\bf \check{X}}_{i})\cdot 
\mathop{\rm Y}%
\nolimits_{k_{i}^{\prime }}({\bf \check{Z}})\right) \left( 
\mathop{\rm Y}%
\nolimits_{k_{i}^{\prime \prime }}({\bf \check{Z}})\cdot 
\mathop{\rm Y}%
\nolimits_{k_{i}^{\prime \prime }}({\bf \check{X}}_{i}^{\prime })\right) 
\nonumber \\
&=&\sum_{\left( l_{1},l_{2}\right) L;(l_{1}^{\prime },l_{2}^{\prime
})L^{\prime };L^{\prime \prime }}{\cal P}_{\{k_{1},k_{2},k_{1}^{\prime
},k_{2}^{\prime },k_{1}^{\prime \prime },k_{2}^{\prime \prime }\};\{\left(
l_{1},l_{2}\right) L;(l_{1}^{\prime },l_{2}^{\prime })L^{\prime };L^{\prime
\prime }\}}  \nonumber \\
&&\left( \left\{ 
\mathop{\rm Y}%
\nolimits_{l_{1}}({\bf \check{X}}_{1})\times 
\mathop{\rm Y}%
\nolimits_{l_{2}}({\bf \check{X}}_{2})\right\} _{L}\cdot \left\{ \left\{ 
\mathop{\rm Y}%
\nolimits_{l_{1}^{\prime }}({\bf \check{X}}_{1}^{\prime })\times 
\mathop{\rm Y}%
\nolimits_{l_{2}^{\prime }}({\bf \check{X}}_{2}^{\prime })\right\}
_{L^{\prime }}\times 
\mathop{\rm Y}%
{}_{L^{\prime \prime }}({\bf \check{Z}})\right\} _{L}\right)
\label{eq:AntisymProdHarmonics}
\end{eqnarray}
The intermediate (but redundant) index $L^{\prime \prime }$ is connected to
the integration variable ${\bf Z}$ of the integral transformation (\ref
{eq:IntegralTransform}). Because of the orthogonality between spherical
harmonics and of the scalar character of the potential operator $L$ and $%
L^{\prime }$ will be equal, and $L^{\prime \prime }=0$ after integration
over ${\bf Z}$: we therefore anticipate by simplifying (\ref
{eq:AntisymProdHarmonics}) to 
\begin{eqnarray}
P_{k_{1},k_{2},k_{1}^{\prime },k_{2}^{\prime },k_{1}^{\prime \prime
},k_{2}^{\prime \prime }} &=&\sum_{\left( l_{1},l_{2}\right)
L;(l_{1}^{\prime },l_{2}^{\prime })L}{\cal P}_{\{k_{1},k_{2},k_{1}^{\prime
},k_{2}^{\prime },k_{1}^{\prime \prime },k_{2}^{\prime \prime }\};\{\left(
l_{1},l_{2}\right) L;(l_{1}^{\prime },l_{2}^{\prime })L;0\}}\frac{1}{\sqrt{%
4\pi }}  \nonumber \\
&&\left( \left\{ 
\mathop{\rm Y}%
\nolimits_{l_{1}}({\bf \check{X}}_{1})\times 
\mathop{\rm Y}%
\nolimits_{l_{2}}({\bf \check{X}}_{2})\right\} _{L}\times \left\{ 
\mathop{\rm Y}%
\nolimits_{l_{1}^{\prime }}({\bf \check{X}}_{1}^{\prime })\times 
\mathop{\rm Y}%
\nolimits_{l_{2}^{\prime }}({\bf \check{X}}_{2}^{\prime })\right\}
_{L}\right)
\end{eqnarray}
The expansion coefficients are then easily shown to be 
\begin{eqnarray}
&&{\cal P}_{\{k_{1},k_{2},k_{1}^{\prime },k_{2}^{\prime },k_{1}^{\prime
\prime },k_{2}^{\prime \prime }\};\{\left( l_{1},l_{2}\right)
L;(l_{1}^{\prime },l_{2}^{\prime })L;0\}}=\sum_{k}(-1)^{l_{1}+l_{1}^{\prime
}+k+L}\left\{ 
\begin{array}{ccc}
l_{2}^{\prime } & l_{2} & k \\ 
l_{1} & l_{1}^{\prime } & L
\end{array}
\right\} C_{k0k0}^{00}  \nonumber \\
&&\prod_{i=1}^{2}\left( -1\right) ^{k_{1}}\left( 2k_{i}+1\right) \left(
2k_{i}^{\prime }+1\right) \left( 2k_{i}^{\prime \prime }+1\right) \left\{ 
\begin{array}{ccc}
k_{i}^{\prime \prime } & l_{i}^{\prime } & k_{i} \\ 
l_{i} & k_{i}^{\prime } & k
\end{array}
\right\} C_{k_{i}^{\prime }0k_{i}0}^{l_{i}0}C_{k_{i}^{\prime \prime
}0k_{i}0}^{l_{i}^{\prime }0}C_{k_{i}^{\prime }0k_{i}^{\prime \prime }0}^{k0}
\end{eqnarray}

Under the anticipative assumptions $L=L^{\prime }$ and $L^{\prime \prime }=0$%
, the integration over\ the angles ${\bf \check{Z}}$ in (\ref
{eq:PotGenFunIntZfinal}) is now trivial. The remaining integration over $Z$
is easily done after substituting the power expansions for the Bessel
functions $%
\mathop{\rm j}%
_{l}(x)$ and $%
\mathop{\rm i}%
_{l}(x)$. The final result provides a tractable though very bulky result for 
${\cal V}_{\left( l_{1}l_{2}\right) LM;(l_{1}l_{2})LM}\left( \epsilon
,S;\epsilon ^{\prime },S^{\prime }\right) $\ that is not reproduced here, as
it carries no further additional information.

Again, by using the standard procedures such as differentiation or
recurrence relations one obtains the effective matrix elements. In
particular the form of (\ref{eq:MatrElemVbyDiffer}) remains valid, though
the analytic differentiation preferably should be performed within an
algebraic package such as Mathematica or Maple due to the bulkiness of the
formulae.

\subsection{Matrix elements for the Coulomb potential}

The matrix elements of the Coulomb potential can now be most easily obtained
from the Gaussian results. We consider again the Gaussian integral
representation 
\begin{equation}
\frac{1}{\left| {\bf r}_{i}-{\bf r}_{j}\right| }=\frac{2}{\sqrt{\pi }}%
\int_{0}^{\infty }dx\exp \{-x^{2}\,({\bf r}_{i}-{\bf r}_{j})^{2}\}
\end{equation}
We use (\ref{eq:PotGenFunIntZfinal}), replace $a^{2}$ by $1/x^{2}$, and
introduce a new integration variable $t=\gamma /(1-\gamma )$ where $\gamma
=x^{2}b^{2}$, to obtain symbolically 
\begin{eqnarray}
&&V_{C}\left( \epsilon {\bf ,}S{\bf ,\check{X}}_{1}{\bf ,\check{X}}_{{\bf 2}%
};\epsilon ^{\prime }{\bf ,}S^{\prime }{\bf ,\check{X}}_{1}^{\prime }{\bf ,%
\check{X}}_{{\bf 2}}^{\prime }\right)  \nonumber \\
&=&\frac{2}{\sqrt{\pi }}\int_{0}^{\infty }dx\ V\left( \epsilon {\bf ,}S{\bf ,%
\check{X}}_{1}{\bf ,\check{X}}_{{\bf 2}};\epsilon ^{\prime }{\bf ,}S^{\prime
}{\bf ,\check{X}}_{1}^{\prime }{\bf ,\check{X}}_{{\bf 2}}^{\prime }\right) 
\nonumber \\
&=&\frac{2}{b\,\sqrt{\pi }}\int_{0}^{1}\frac{dt}{(1-t)^{\frac{3}{2}}\,t^{%
\frac{1}{2}}}V\left( \epsilon {\bf ,}S{\bf ,\check{X}}_{1}{\bf ,\check{X}}_{%
{\bf 2}};\epsilon ^{\prime }{\bf ,}S^{\prime }{\bf ,\check{X}}_{1}^{\prime }%
{\bf ,\check{X}}_{{\bf 2}}^{\prime }\right)  \label{eq:CoulAsGaussTransfrm}
\end{eqnarray}
Mutatis mutandis we apply the integral transformation directly to the
generating matrix elements ${\cal V}_{\left( l_{1}l_{2}\right)
LM;(l_{1}^{\prime }l_{2}^{\prime })LM}\left( \epsilon ,S;\epsilon ^{\prime
},S^{\prime }\right) $%
\begin{eqnarray}
&&{\cal V}_{\left( l_{1}l_{2}\right) LM;(l_{1}^{\prime }l_{2}^{\prime
})LM}^{C}\left( \epsilon ,S;\epsilon ^{\prime },S^{\prime }\right)  \nonumber
\\
&=&\frac{2}{\sqrt{\pi }}\int_{0}^{\infty }dx\ {\cal V}_{\left(
l_{1}l_{2}\right) LM;(l_{1}^{\prime }l_{2}^{\prime })LM}\left( \epsilon
,S;\epsilon ^{\prime },S^{\prime }\right)  \nonumber \\
&=&\frac{2}{b\,\sqrt{\pi }}\int_{0}^{1}\frac{dt}{(1-t)^{\frac{3}{2}}\,t^{%
\frac{1}{2}}}{\cal V}_{\left( l_{1}l_{2}\right) LM;(l_{1}^{\prime
}l_{2}^{\prime })LM}\left( \epsilon ,S;\epsilon ^{\prime },S^{\prime }\right)
\label{eq:CoulRedAsGaussTransform}
\end{eqnarray}
This however leads to a very intricate evaluation of the integral, and thus
for the reduced generating matrix element of the Coulomb potential.

A better procedure is to generate the quantum numbers $K$ and $K^{\prime }$
first, by differentiation of ${\cal V}_{\left( l_{1}l_{2}\right)
LM;(l_{1}^{\prime }l_{2}^{\prime })LM}\left( \epsilon ,S;\epsilon ^{\prime
},S^{\prime }\right) $ on $S$ and $S^{\prime }$, then setting $S=S^{\prime
}=0$. This leads to reduced generating functions for each $K$ and $K^{\prime
}$ now only dependent on $\epsilon $ and $\epsilon ^{\prime }$, which are of
a much simpler form and allow for an analytic integration in (\ref
{eq:CoulRedAsGaussTransform}). The further derivation on matrix elements for 
$n$ and $n^{\prime }$ is then straightforward.

\section{Conclusion}

In this paper we presented a framework for a microscopic three-cluster model
within the Algebraic Model for scattering. It was shown that it is possible
to obtain matrix elements for fully antisymmetrized three-cluster
configurations, as well as a proper description for the the three-cluster
continuum in terms of a hyperspherical description. The corresponding AM
equations in a multichannel description were also introduced. In the current
work the individual clusters were limited to contain only s-orbitals, thus
reducing the mass of clussters to that of a four-nucleon system. The latter
restriction is however not a fundamental one, and was taken to restrict the
analytical and calculational burden.

In order to prove the validity and feasibility of the current model we will
apply it to two specific three-cluster configurations believed to be of
importance to astrophysical physiscs, $\alpha +n+n$ for $^{6}$He, and $%
\alpha +p+p$ for $^{6}$Be. These results appear in part II of this paper.

\section{Acknowledgments}

This work was partially supported by an INTAS
Grant-(``INTAS-93-755-extension). One of the authors (V. V.) gratefully
acknowledges the University of Antwerp (RUCA) for a ``RAFO-gastprofessoraat
1998-1999'' and the kind hospitality of the members of the research group
``Computational Quantum Physics'' of the Department of ``Mathematics and
Computer Sciences'', University of Antwerp, RUCA, Belgium.

\end{document}